\documentclass[11pt]{article}%
\usepackage[bookmarks, linkcolor=blue, urlcolor=blue]{hyperref}
\usepackage{xr-hyper}
\usepackage{amsmath}
\usepackage{amsthm}
\usepackage{amssymb}
\usepackage{amsfonts}
\usepackage{mathtools}
\usepackage{float}
\usepackage{pgffor}
\usepackage{pgf}
\usepackage{tikz,pgfplots}
\usepackage{graphicx}
\usepackage{comment}
\usepackage{caption}
\usepackage{adjustbox}
\usepackage[authoryear]{natbib}
\usepackage{dcolumn}
\usepackage{rotating}
\usepackage[flushleft]{threeparttable}
\usepackage{pdflscape}
\usepackage{afterpage}
\usepackage[pass]{geometry}
\usepackage{setspace}
\usepackage{bigstrut}
\usepackage{footnote}
\usepackage{calc}
\usepackage{accents}%
\providecommand{\U}[1]{\protect\rule{.1in}{.1in}}
\newcolumntype{L}{D{.}{.}{2,2}}

\definecolor{MyDarkBlue}{rgb}{0.1,0.2,0.65}
\hypersetup{
	colorlinks   = true,
	citecolor    = MyDarkBlue
}
\specialcomment{mycomment} {\begingroup\color{gray}}{\endgroup}
\excludecomment{mycomment}
\usepgflibrary{plothandlers}

\newtheorem{proposition}{Proposition}[section]
\newtheorem{lemma}[proposition]{Lemma}

\newenvironment{pff}[1][Proof]{\vspace{1ex}{\noindent\textbf{#1.} }\hspace{.1em}}
{\hfill\qed\vspace{1ex}}
\newtheorem{theo}{Theorem}
\newtheorem{assumption}{Assumption}
\newtheorem{condition}{Condition}
\newtheoremstyle{examm}
{\topsep}{\topsep}
{\upshape}
{}
{\bfseries}
{.}
{ }
{}
\theoremstyle{examm}
\newtheorem{example2}{Example}
\newtheorem{rem2}{Remark}
\theoremstyle{definition}

\theoremstyle{remark}

\numberwithin{equation}{section}
\newcommand*\diff{\mathop{}\!\mathrm{d}}
\DeclareMathOperator*{\argmax}{argmax}
\let\originalleft\left
\let\originalright\right
\renewcommand{\left}{\mathopen{}\mathclose\bgroup\originalleft}
\renewcommand{\right}{\aftergroup\egroup\originalright}

\newcommand\betatilde{\stackrel{\sim}{\smash{\beta}\rule{0pt}{1.13ex}}}

\newcommand\lambdatilde{\stackrel{\sim}{\smash{\lambda}\rule{0pt}{1.13ex}}}

\newcommand\sigmatilde{\stackrel{\sim}{\smash{\sigma}\rule{0pt}{0.63ex}}}

\newcommand\sbullet[1][.4]{\mathbin{\vcenter{\hbox{\scalebox{#1}{$\bullet$}}}}}
\begin{document}

\begin{center}
\renewcommand{\thefootnote}{\fnsymbol{footnote}} \begin{savenotes}
		{\Large \textbf{Non-Identifiability in Network Autoregressions}}\\
		\vspace{0.7cm}Federico Martellosio\footnote{School of Economics, University of Surrey, f.martellosio@surrey.ac.uk}\\
		\bigskip
		\hspace{.1cm}June 1, 2022
	\end{savenotes}
\setcounter{footnote}{0}
\end{center}

\bigskip

\noindent\textbf{Abstract}

\noindent We study identifiability of the parameters in autoregressions defined on a
network. Most identification conditions that are available for these models
either rely on the network being observed repeatedly, are only sufficient, or
require strong distributional assumptions. This paper derives conditions that
apply even when the individuals composing the network are observed only once,
are necessary and sufficient for identification, and require weak
distributional assumptions. We find that the model parameters are generically,
in the measure theoretic sense, identified even without repeated observations,
and analyze the combinations of the interaction matrix and the regressor
matrix causing identification failures. This is done both in the original
model and after certain transformations in the sample space, the latter case
being relevant, for example, in some fixed effects specifications.

\bigskip

\bigskip

\vspace*{0.2cm} \noindent\textit{Keywords}: fixed effects, invariance,
networks, quasi maximum likelihood estimation.\newline\noindent\textit{JEL}
\textit{Classification}: C12, C21.

\section{Introduction}

In a wide range of empirical settings, data are available for an outcome
variable and some covariates for each of the nodes of a network, as well as
some measure of the pairwise interaction between the nodes. Autoregressive
processes offer a simple way to study how the covariates affect the outcome
variable, taking into account the network interaction. Models of this type can
be traced back at least to \cite{whittle1954}, and have since proved useful in
many applications, across many scientific fields. In economics, and the social
sciences more generally, they are currently particularly popular in the
analysis of peer effects and social networks. The models are known as
simultaneous autoregressions in the statistics literature
\citep[e.g.,][]{Cressie1993}, spatial autoregressions in the econometrics
literature \citep[e.g.,][]{LeSagePace2009}, are closely related to
linear-in-means models \citep[e.g.,][]{Manski93}, and have important
connections to linear structural equation models \citep[e.g.,][]{Drton2011}.
To emphasize their wide applicability, we refer to them as network autoregressions.

This paper is concerned with identifiability of the parameters in a network
autoregression. We employ a classical notion of identifiability, according to
which parameters are identified if they are uniquely recovered from the
distribution of the observables. Lack of identification has, of course,
serious consequences for inference. For example, identification is necessary
for the limiting objective function of an extremum estimator to be uniquely
maximized at the true value of the parameter, which is a standard condition
for consistency \citep[see, e.g.,][]{NeweyMcFadden94}. In addition, inference
is expected to be difficult near the cases in which identification fails.
Given how fundamental the problem of identification is, and given that
establishing consistency of some extremum estimator is sufficient for
identification, it is not surprising that there is a vast amount of work that
is relevant for the present study. We mention in particular two very
influential papers: \cite{Lee2004} and \cite{Bramoulle2009}. \cite{Lee2004},
in a rigorous analysis of the asymptotic properties of the quasi maximum
likelihood estimator based on the Gaussian distribution, provides conditions
that are sufficient for consistency and hence for identifiability.
\cite{Bramoulle2009} investigates identifiability by looking at the mapping
from the reduced form parameters to the structural parameters, an approach
that has become standard in the social network literature.

The present paper studies identifiability directly from the first moment, or
the first two moments, of the outcome variable. Compared to the approach via
reduced form parameters, identification from moments does not require the
nodes of the network being observed over multiple instances (e.g., over time).
We show that identification from the first moment is generally possible, and
characterize the cases when it is impossible. One class of cases when
identifiability from the first moment is impossible is particularly relevant
in fixed effects models (for example, the classical linear-in-means model with
group fixed effects belongs to this class of cases). In that class,
non-identifiability from the first moment is linked to the impossibility of
invariant inference; more precisely, the parameters cannot be identified from
any statistic that is invariant with respect to a certain group of
transformations under which the model itself is invariant. This type of
non-identifiability occurs despite the fact that the parameters may be
identifiable from the second moment of the outcome variable. Hence,
identifiability from the second moment is of very questionable value in these
case, because it can only lead to non-invariant inference.

\begin{mycomment}
we aim to understand what combinations of the interaction matrix $W$ and the
regressor matrix $X$ lead to a failure of identification. ..........
The present paper departs from previous studies in two main ways.
[[[this bit was in main text, move it to intro: Also, Condition \ref{assum id}
will be analyzed in detail in Sections \ref{sec inv} and \ref{sec lik}. In
particular, it will become clear in Section \ref{sec inv} that a failure of
Condition \ref{assum id} implies a specific type of non-identifiability.]]]
\end{mycomment}

Section \ref{sec SLM} sets out the framework. Section \ref{sec ident} studies
identifiability from the first and second moments of the outcome variable, and
Section \ref{sec inv} discusses identifiability after reduction by invariance.
Section \ref{Sec simul} reports simulation evidence on the consequences of
being close to non-identifiability. Section \ref{sec concl} briefly concludes.
The appendices contain additional material and all proofs. Throughout the
paper the results are illustrated by means of several examples.

\textit{Notation}. Matrices are denoted by capital letters, vectors and
scalars by lowercase letters. We reserve bold letters for random quantities
(scalars, vectors, or matrices), so, for example, $\boldsymbol{y}$ denotes a
random vector and $y$ a realization of $\boldsymbol{y}$. Throughout the paper,
$\iota_{n}$ denotes the $n\times1$ vector of ones, $\mathbb{R}^{n\times m}$
denotes the set of real $n\times m$ matrices, $\operatorname{col}(A)$ denotes
the column space of a matrix $A$, $M_{A}\coloneqq I_{n}-A(A^{\prime}%
A)^{-1}A^{\prime}$ for a full column rank matrix $A$, $\mu_{\mathbb{R}^{n}}$
denotes the Lebesgue measure on $\mathbb{R}^{n}$, \textquotedblleft
a.s.\textquotedblright\ stands for almost surely, with respect to
$\mu_{\mathbb{R}^{n}}$, and $A\oplus B$ denotes the direct sum of the matrices
$A$ and $B$ (that is, if $A$ is $n\times m$ and $B$ is $p\times q$, $A\oplus
B$ is the $(n+p)\times(m+q)$ block diagonal matrix with $A$ as top diagonal
block and $B$ as bottom diagonal block).

\section{\label{sec SLM}The model}

The model of interest is the \textit{network autoregression}
\begin{equation}
\boldsymbol{y}=\lambda W\boldsymbol{y}+X\beta+\sigma\boldsymbol{\varepsilon},
\label{SLM}%
\end{equation}
where $\boldsymbol{y}$ is the $n\times1$ vector of outcomes, $\lambda$ is a
scalar parameter, $W$ is an interaction matrix, $X$ is an $n\times k$ matrix
of regressors with full column rank and with $k\leq n-2$, $\beta\in
\mathbb{R}^{k}$, $\sigma$ is a positive scale parameter, and
$\boldsymbol{\varepsilon}$ is an unobservable $n\times1$ random vector with
$\mathrm{E}(\boldsymbol{\varepsilon})=0$. The matrices $W$ and $X$ are assumed
to be nonstochastic and known as, for instance, in \cite{Lee2004}%
.\footnote{\label{footnote stoch}At the cost of some additional complexity,
one could, alternatively, take $W$ and/or $X$ to be stochastic, and condition
on $W$ and/or $X$ under suitable exogeneity assumptions
\citep[see, e.g.,][]{Bramoulle2009,Gupta2019}. In that case the assumption
$\mathrm{E}(\boldsymbol{\varepsilon})=0$ would be replaced by $\mathrm{E}%
(\boldsymbol{\varepsilon}\mid W,X)=0$. Allowing for endogeneity of $W$ and/or
$X$ would instead require different methods; see Section \ref{sec concl}.} The
entries of $W$ are supposed to reflect the pairwise interaction between the
observational units; in particular, the $(i,j)$-th entry of $W$ is zero if
unit $j$ is not deemed to be a neighbor of unit $i$. Some of the columns of
$X$ may be spatial lags of some other columns (the spatial lag of a vector $x$
being the vector $Wx$). That is, in the terminology of social networks, we
allow for \textquotedblleft contextual effects\textquotedblright\ or
\textquotedblleft exogenous spillovers\textquotedblright. We assume that
$\lambda$ is such that the model has a unique reduced form, or, in other
words, that $\boldsymbol{y}$ is uniquely determined given $X$\ and
$\boldsymbol{\varepsilon}$.\footnote{\label{footnote incomplete}Identification
analysis when a unique reduced form does not exist would require different
tools; see, e.g., \cite{ChesherRosen2008}.} This requires $S(\lambda
)\coloneqq
I_{n}-\lambda W$ to be nonsingular. We refer to the set $\Lambda_{\mathrm{u}%
}\coloneqq\{\lambda\in\mathbb{R}:\det(S(\lambda))\neq0\}$ as the unrestricted
parameter space for $\lambda$. Note that the values of the (real) parameter
$\lambda$ such that $\det(S(\lambda))=0$ are $\lambda=\omega^{-1}$ for any
nonzero real eigenvalue $\omega$ of $W$, so $\Lambda_{\mathrm{u}}$ is the
whole real line minus a number (less or equal to $n$) of isolated points.

\begin{mycomment}
	If there is no $X$, the model is
	referred to as a pure network autoregression.
\end{mycomment}

When the index set of $\boldsymbol{y}$ has more than one dimension (e.g.,
individuals and time, or individuals and networks), it is often useful to
include in the error term additive unobserved components relative to those
dimensions. In that case, we take a fixed effects approach and treat the
unobserved effects as parameters to be estimated. Accordingly, for inferential
purposes, we incorporate the fixed effects into $\beta$ and the corresponding
dummy variables into $X$. Two examples of fixed effects specifications that
can be nested into the general model (\ref{SLM}) are given next.

\begin{mycomment}
(so the dimension of $\beta$ may be increasing with the sample size)
\end{mycomment}

\begin{example2}
\label{exa panel}(\textit{Panel data model}) There are $N$ individuals,
followed over $T$ time periods. Let $W_{t}$ be an $N\times N$ matrix
describing the interaction between individuals at time $t$, and $\widetilde{X}%
$ an $NT\times\tilde{k}$ regressor matrix. A panel data version of the network
autoregression (\ref{SLM}) is given by $\boldsymbol{y}_{it}=\lambda\sum
_{ij}(W_{t})_{ij}\boldsymbol{y}_{jt}+\widetilde{x}_{it}^{\prime}\tilde{\beta
}+\boldsymbol{u}_{it}$, for $i=1,\ldots,N$ and $t=1,\ldots,T$, where
$(W_{t})_{ij}$ are the entries of $W_{t}$, and $\widetilde{x}_{it}^{\prime}$
are the $\tilde{k}\times1$ rows of $\widetilde{X}.$ The error $\boldsymbol{u}%
_{it}$ is decomposed into $\boldsymbol{c}_{i}+\sigma\boldsymbol{\varepsilon
}_{it}$ (one-way model) or $\boldsymbol{c}_{i}+\boldsymbol{\alpha}_{t}%
+\sigma\boldsymbol{\varepsilon}_{it}$ (two-way model), where $\boldsymbol{c}%
_{i}$ and $\boldsymbol{\alpha}_{t}$ are, respectively, individual specific
effects and time specific effects, and $\boldsymbol{\varepsilon}_{it}$ is an
idiosyncratic error. Following a fixed effects approach (i.e., treating the
random components $\boldsymbol{c}_{i}$ and $\boldsymbol{\alpha}_{t}$ as
parameters to be estimated), the model can be written in the notation of
equation (\ref{SLM}), with $W=\bigoplus_{t=1}^{T}W_{t}$, and, for the two-way
model, $X=(\widetilde{X},\iota_{T}\otimes I_{N},I_{T}\otimes\iota_{N})$ and
$\beta=(\tilde{\beta}^{\prime},c^{\prime},\alpha^{\prime})^{\prime}$, where
$c$ and $\alpha$ are the vectors with entries $c_{i}$ and $\alpha_{t}$,
respectively.\footnote{Obviously, for identification of $\beta$, one column of
the matrix $(\iota_{T}\otimes I_{N},I_{T}\otimes\iota_{N})$ should be omitted
from $X$, or some normalization should be imposed on the fixed effects, and no
regressor should be constant over time or over individuals.} In most
applications, $W_{t}$ is taken to be time invariant, say $W_{t}=W^{\ast}$ for
all $t=1,\ldots,T$, so that $W=I_{T}\otimes W^{\ast}$.\hfill\qed

\end{example2}

\begin{example2}
\label{exa network}(\textit{Network fixed effects}) There are $R$ networks,
with network $r$ having $m_{r}$ individuals. The model is
\begin{equation}
\boldsymbol{y}_{r}=\lambda W_{r}\boldsymbol{y}_{r}+\widetilde{X}_{r}%
\tilde{\beta}+\boldsymbol{\alpha}_{r}\iota_{m_{r}}+\sigma
\boldsymbol{\varepsilon}_{r},\text{ }r=1,\ldots,R, \label{network model}%
\end{equation}
where $W_{r}$ is the $m_{r}\times m_{r}$ interaction matrix of network $r$,
$\widetilde{X}_{r}$ is the $m_{r}\times\tilde{k}$ regressor matrix of network
$r$, $\tilde{\beta}$ is a $\tilde{k}\times1$ parameter, and
$\boldsymbol{\alpha}_{r}$ is a network fixed effect. Stacking the equations in
(\ref{network model}) vertically, and following a fixed effects approach, the
model can be written in the notation of equation (\ref{SLM}), with
$\boldsymbol{y}=(\boldsymbol{y}_{1}^{\prime},\ldots,\boldsymbol{y}_{R}%
^{\prime})^{\prime}$, $W=\bigoplus_{r=1}^{R}W_{r}$, $\beta=(\tilde{\beta
}^{\prime},\alpha^{\prime})^{\prime}$, $\boldsymbol{\varepsilon}%
=(\boldsymbol{\varepsilon}_{1}^{\prime},\ldots,\boldsymbol{\varepsilon}%
_{R}^{\prime})^{\prime}$, and $X=(\widetilde{X},\bigoplus_{r=1}^{R}%
\iota_{m_{r}})$, where $\widetilde{X}\coloneqq(\widetilde{X}_{1}^{\prime
},\ldots,\widetilde{X}_{R}^{\prime})^{\prime}$.\hfill\qed

\end{example2}

In the rest of the paper, unobserved effects are always treated as parameters.
Two specific network autoregressions that will be used to illustrate our
results are as follows.

\begin{example2}
\label{exa GI}(\textit{Group Interaction model}) A particular case of model
(\ref{network model}), which we refer to as the Group Interaction model, is
when all members of a group interact homogeneously, that is, $W_{r}=\frac
{1}{m_{r}-1}(\iota_{m_{r}}\iota_{m_{r}}^{\prime}-I_{m_{r}})\eqqcolon B_{m_{r}%
}$, for $r=1,\ldots,R$. Following \cite{Manski93}, this specific structure has
played a central role in the literature on peer effects. We say that the Group
Interaction model is \textit{balanced} if all group sizes $m_{r}$ are the
same. In that case, letting $m$ denote the common group size, $W=I_{R}\otimes
B_{m}$.\hfill\qed

\end{example2}

\begin{example2}
\label{exa CBG}(\textit{Complete Bipartite model}) In a complete bipartite
network the $n$ observational units are partitioned into two groups, of sizes
$p$ and $q$ say, with all units within a group interacting with all in the
other group, but with none in their own group \citep[e.g.,][]{wasserman94}. In
economics, such a structure arises commonly when modeling two-sided markets,
before any specific matching between the two groups has taken place or when
information about matchings is not available. The two groups could be, for
instance, buyers and sellers, with each seller interacting with all buyers,
and each buyer interacting with all sellers. For $p=1$ or $q=1$ this
corresponds to the network known as a \textit{star}. The adjacency matrix of a
complete bipartite network is
\[
A\coloneqq\left(
\begin{array}
[c]{cc}%
0_{pp} & \iota_{p}\iota_{q}^{\prime}\\
\iota_{q}\iota_{p}^{\prime} & 0_{qq}%
\end{array}
\right)  .
\]
The associated row-normalized interaction matrix is\footnote{For an entrywise
nonnegative matrix $B$ having all row-sums different from zero, the
row-normalized version of $B$ is obtained by dividing each entry of $B$ by the
corresponding row-sum, and is therefore a row-stochastic matrix.}
\begin{equation}
W=\left(
\begin{array}
[c]{cc}%
0_{pp} & \frac{1}{q}\iota_{p}\iota_{q}^{\prime}\\
\frac{1}{p}\iota_{q}\iota_{p}^{\prime} & 0_{qq}%
\end{array}
\right)  . \label{row-std CBG}%
\end{equation}
Alternatively, $A$ can be rescaled by its largest eigenvalue, yielding the
symmetric interaction matrix
\begin{equation}
W=\frac{1}{\sqrt{pq}}A. \label{symm CBG}%
\end{equation}
We refer to the network autoregressions with interaction matrix
(\ref{row-std CBG}) or (\ref{symm CBG}), as, respectively, the
\textit{row-normalized} Complete Bipartite model and the \textit{symmetric}
Complete Bipartite model. \hfill\qed

\end{example2}

\section{\label{sec ident}Identifiability}

This section explores identifiability of the parameters of a network
autoregression in two cases. First, Section \ref{sec id first} discusses what
can be identified when no probabilistic assumptions beyond the maintained
assumption $\mathrm{E}(\boldsymbol{\varepsilon})=0$ are imposed on the model.
Then, Section \ref{sec id II moment} considers adding assumptions on the
second moment of $\boldsymbol{\varepsilon}$. Connections to the literature are
discussed in Section \ref{sec literat}.

We employ a classical notion of global identifiability, according to which a
parameter is said to be identified if it can be uniquely recovered from the
distribution of the observables
\citep[see, e.g.,][]{Koopmans1950, Rothenberg1971, Matzkin2007}. The precise
definitions we need are as follows. Consider a statistical model, defined as a
family of distributions $\left\{  P_{\theta}:\theta\in\Theta\subseteq
\mathbb{R}^{p}\right\}  $ for some observable random vector on a certain
sample space. A particular value $\theta_{1}\in\widetilde{\Theta}%
\subseteq\Theta$ of $\theta$ is said to be \textit{identified} (from the
distribution $P_{\theta}$) on a set $\widetilde{\Theta}$ if there is no other
$\theta_{2}\in\widetilde{\Theta}$ such that $P_{\theta_{1}}=P_{\theta_{2}}$.
If all values of $\theta$ in $\widetilde{\Theta}$ are identified on
$\widetilde{\Theta}$, we say that the parameter $\theta$ is identified on
$\widetilde{\Theta}$ (if the set $\widetilde{\Theta}$ is the whole parameter
space $\Theta$, one often simply says that the model is identified). If all
values of $\theta$ in $\widetilde{\Theta}$ except for those in a
$\mu_{\mathbb{R}^{p}}$-null set are identified on $\widetilde{\Theta}$, we say
that the parameter $\theta$ is \textit{generically} \textit{identified} on
$\widetilde{\Theta}$. Identifiability can also be applied to functions of the
parameter $\theta$, so that a function $f(\theta)$ is identified if it can be
uniquely recovered from $P_{\theta}$. Formally, the function $f(\theta)$ is
said to be identified on $\widetilde{\Theta}$ if $f(\theta_{1})\neq
f(\theta_{2})$ implies $P_{\theta_{1}}\neq P_{\theta_{2}}$ for any $\theta
_{1},\theta_{2}\in\widetilde{\Theta}$. The function $f(\theta)$ may extract a
component of $\theta$. Note that a sufficient condition for a function
$g(\theta)$ to be identified (on a set $\widetilde{\Theta}$) is that it can be
recovered uniquely from a function $f(\theta)$ that is identified.\footnote{To
see this, take arbitrary $\theta_{1},\theta_{2}$ (in $\widetilde{\Theta}$)
such that $g(\theta_{1})\neq g(\theta_{2})$, and assume $g(\theta)$ can be
recovered uniquely from $f(\theta)$, so that $f(\theta_{1})\neq f(\theta_{2}%
)$. Then $P_{\theta_{1}}\neq P_{\theta_{2}}$, because $f(\theta)$ is
identified, which shows that $g(\theta)$ is identified.} Moments of
$P_{\theta}$ (for example the mean, the variance matrix) are certainly
identified functions of $\theta$ (because different moments imply different
distributions), and so a sufficient condition for a function $g(\theta)$ to be
identified (on a set $\widetilde{\Theta}$) is that it can be recovered
uniquely from a moment of $P_{\theta}$; in this case we say that $g(\theta)$
is \textit{identified from a moment}.

Note that identification from a moment is different from the concept, relevant
in GMM estimation, of identification from a moment condition. Indeed, the
concept of identification this paper refers to is detached from the choice of
an estimator, unless one assumes that the distribution $P_{\theta}$ is known
(up to $\theta$), in which case identification is equivalent to identification
based on the (correctly specified) likelihood \citep{Rothenberg1971}.
Identification based on an extremum estimator is sufficient, but in general
not necessary, for identification \citep[e.g.,][]{NeweyMcFadden94}.

\subsection{\label{sec id first}Identifiability from first moment}

We start by studying generic identification, according to the definition just
given, of $\lambda$ and $\beta$ when no distributional assumptions beyond
$\mathrm{E}(\boldsymbol{\varepsilon})=0$ are imposed on the model.\footnote{Of
course, the scale parameter $\sigma$, as well as any parameters that might be
used to parametrize the variance of $\boldsymbol{\varepsilon}$, cannot be
identified assuming only $\mathrm{E}(\boldsymbol{\varepsilon})=0$.
Identification of $\sigma$ requires only very weak conditions (or a
normalization) on the variance of $\boldsymbol{\varepsilon}$, whereas
identifiability of any parameters affecting the variance of
$\boldsymbol{\varepsilon}$ will depend on the particular parametric
specification.}

\begin{proposition}
\label{lemma identif mean}In the network autoregression (\ref{SLM}),

\begin{enumerate}
\item[(i)] if $\mathrm{rank}(X,WX)>k$, the parameter $(\lambda,\beta)$ is
generically identified on $\Lambda_{\mathrm{u}}\times\mathbb{R}^{k}$;

\item[(ii)] if $\mathrm{rank}(X,WX)=k$, no value of the parameter
$(\lambda,\beta)$ is identified on $\Lambda_{\mathrm{u}}\times\mathbb{R}^{k}$.
\end{enumerate}
\end{proposition}

\begin{mycomment}
	Needless to say, the crucial assumption here
is that $\mathrm{E}(\boldsymbol{\varepsilon})$ does not depend on $\lambda$ and $\beta$.
Also, recall that we are assuming that $W$ and $X$ are nonstochastic;
otherwise, the assumption $\mathrm{E}(\boldsymbol{\varepsilon})=0$ should be replaced by
$\mathrm{E}(\boldsymbol{\varepsilon}|W,X)=0$.
\end{mycomment}

\begin{mycomment}
	In a pure model, $\mathrm{E}(\boldsymbol{y})$ is zero and hence cannot identify
any parameters.
\end{mycomment}

\begin{mycomment}
	maybe just state in the lemma conditions for param to be identified from first
	moment. then connect to identifiability afterwards, since identifiability of
	the moments may not be straightforward; see Goldsmith Imbens 2013 (for identif
	from second moment see de paula review, p 279). BUT\ WAIT, it is actually
	straightforward. If $\mathrm{rank}(X,WX)>k$ then $S^{-1}(\lambda)X\beta
	=S^{-1}(\tilde{\lambda})X\tilde{\beta}$ implies $\left(  \lambda,\beta\right)
	=(\tilde{\lambda},\tilde{\beta})$ for any $\left(  \lambda,\beta\right)
	,(\tilde{\lambda},\tilde{\beta})\in\Lambda_{\mathrm{u}}\times\mathbb{R}^{k}$. That is,
	if $\mathrm{rank}(X,WX)>k$ $\left(  \lambda,\beta\right)  $ is identifiable
\end{mycomment}

Proposition \ref{lemma identif mean} says that the parameters $\lambda$ and
$\beta$ are generically identified (from the first moment of $\boldsymbol{y}$)
if the matrices $X$ and $W$ are such that $\mathrm{rank}(X,WX)>k$. Conversely,
if $\mathrm{rank}(X,WX)=k$, $\lambda$ and $\beta$ cannot be identified, and
hence consistently estimated, without distributional assumptions beyond
$\mathrm{E}(\boldsymbol{\varepsilon})=0$. A manifestation of this is that the
2SLS estimators of \cite{Kelejian98} and \cite{Lee2003}, which are based on
the specification of the first moment only of $\boldsymbol{y}$, are not
defined if $\mathrm{rank}(X,WX)=k$.\footnote{The 2SLS estimator of
\cite{Kelejian98} uses as instruments for the $k+1$ variables $(X,Wy)$ a
subset of the columns of the linearly independent columns of $(X,WX,W^{2}%
X,\ldots)$. But when $\mathrm{rank}(X,WX)=k$, there are at most $k$ such
linearly independent columns.} Note that if $\lambda$ is (generically)
identified on $\Lambda_{\mathrm{u}}$ then it is also (generically) identified
on any subset of $\Lambda_{\mathrm{u}}$.

The condition $\mathrm{rank}(X,WX)=k$ is trivially satisfied when $k=0$; in
that case, $\mathrm{E}(\boldsymbol{y})$ is zero and therefore cannot identify
any parameter. When $k>0$, the condition is typically very
strong,\footnote{\label{footnote leb}Indeed, for any given $W$ that is not a
scalar multiple of the identity matrix, the set of (full rank) $n\times k$
matrices $X$ such that $\mathrm{rank}(X,WX)=k$ is a $\mu_{\mathbb{R}^{n\times
k}}$-null set (with $\mu_{\mathbb{R}^{n\times k}}$ denoting the Lebesgue
measure on the set of real $n\times k$ matrices). Accordingly, Proposition
\ref{lemma identif mean} could be stated by saying that identification from
the first moment of $\boldsymbol{y}$ is possible for generic parameter values
$(\lambda,\beta)$ and for generic regressor matrices $X$.} but specific
combinations of $W$ and $X$ such that $\mathrm{rank}(X,WX)=k$ may arise in
some cases of interest, particularly in fixed effects models. An important
class of cases when $\mathrm{rank}(X,WX)=k$ is characterized by a failure of
the following condition.

\begin{condition}
\label{assum id}There is no real eigenvalue $\omega$ of $W$ for which
$M_{X}(\omega I_{n}-W)=0.$
\end{condition}

Indeed, $M_{X}(\omega I_{n}-W)=0$ implies $M_{X}WX=0$, which is equivalent to
$\mathrm{rank}(X,WX)=k.$ Thus, by Proposition \ref{lemma identif mean}, any
pair of matrices $X$ and $W$ violating Condition \ref{assum id} gives rise to
a model in which $\lambda$ and $\beta$ cannot be identified from
$\mathrm{E}(\boldsymbol{y})$. A condition equivalent to $M_{X}(\omega
I_{n}-W)=0$ is $\operatorname{col}(\omega I_{n}-W)\subseteq\operatorname{col}%
(X)$. That is, a pair $(X,W)$ causes Condition \ref{assum id} to fail if and
only if $\operatorname{col}(X)$ contains the subspace $\operatorname{col}%
(\omega I_{n}-W)$, for some real eigenvalue $\omega$ of $W$. Also, note that
if, for a given $W$, Condition \ref{assum id} is violated for some $X=X_{1}$,
then it is also violated for $X=(X_{1},X_{2})$, for any $X_{2}$ (such that $X$
is full rank). It is helpful to look at how failures of Condition
\ref{assum id} can arise in the contexts of Examples \ref{exa GI} and
\ref{exa CBG}.

\begin{example2}
\label{exa GI failure}For the matrix $W=I_{R}\otimes B_{m}$ of a Balanced
Group Interaction model (Example \ref{exa GI}), the smallest eigenvalue is
$\omega_{\min}=-\frac{1}{m-1}$, and $\operatorname{col}(\omega_{\min}%
I_{n}-W)=\operatorname{col}(I_{R}\otimes\iota_{m})$. Since $I_{R}\otimes
\iota_{m}$ is the design matrix of the group fixed effects, it follows that
Condition \ref{assum id} is violated if group fixed effects are included in a
Balanced Group Interaction model.\hfill\qed

\end{example2}

\begin{example2}
\label{exa CBG failure}For both the row-normalized Complete Bipartite model
and the symmetric Complete Bipartite model (Example \ref{exa CBG}),
$\operatorname{col}(W)$ is spanned by the vectors $(\iota_{p}^{\prime}%
,0_{q}^{\prime})^{\prime}$ and $(0_{p}^{\prime},\iota_{q}^{\prime})^{\prime}$.
Hence, for both models, Condition \ref{assum id} is violated (for $\omega=0$)
if $\operatorname{col}(X)$ contains $(\iota_{p}^{\prime},0_{q}^{\prime
})^{\prime}$ and $(0_{p}^{\prime},\iota_{q}^{\prime})^{\prime}$. This is the
case if the model contains an intercept for each of the two groups.\hfill
\qed

\end{example2}

Examples \ref{exa GI failure} and \ref{exa CBG failure} give cases in which
Condition \ref{assum id} is violated, and hence $\mathrm{rank}(X,WX)=k$.
Further examples in which Condition \ref{assum id} fails are given in Appendix
\ref{app further}. We now turn our attention to examples in which
$\mathrm{rank}(X,WX)=k$ even though Condition \ref{assum id} is satisfied.

\begin{example2}
\label{exa netw under cond 1}The condition $\mathrm{rank}(X,WX)=k$ is
satisfied in the following instances of model (\ref{network model}), when
there are at least two groups ($R>1$):\footnote{\label{foot Xtilde}In case
(i), Condition \ref{assum id} is satisfied for generic matrices $\widetilde{X}%
_{1},\ldots,\widetilde{X}_{R}$ if the model is unbalanced, and is violated if
the model is balanced (see Example \ref{exa GI}). In cases (ii) and (iii),
Condition \ref{assum id} is satisfied for any $\widetilde{X}$.}

\begin{enumerate}
\item[(i)] A Balanced Group Interaction model with contextual effects
\citep[e.g.,][]{Liu2017}. The model equation is $\boldsymbol{y}_{r}=\lambda
B_{m}\boldsymbol{y}_{r}+\alpha\iota_{m}+\widetilde{X}_{r}\tilde{\beta}%
+B_{m}\widetilde{X}_{r}\delta+\sigma\boldsymbol{\varepsilon}_{r},$ for
$r=1,\ldots,R,$ where $\alpha$ is an intercept and $\widetilde{X}_{r}$ is
$m\times\tilde{k}$, with $0<\tilde{k}<R$. The matrix $X$ is given by
$(\iota_{n},\widetilde{X},(I_{r}\otimes B_{m})\widetilde{X}),$ and has
therefore $k=2\tilde{k}+1$ columns (recall that $\widetilde{X}$ is the
$n\times\tilde{k}$ matrix obtained by stacking the matrices $\widetilde{X}%
_{1},\ldots,\widetilde{X}_{r}$
vertically).\footnote{\label{footnote BGI context}The condition $\mathrm{rank}%
(X,WX)=k$ is satisfied whether the intercept is included in the model or not.
Also, note that in this model, group fixed effects cannot be added, because
$(\widetilde{X},(I_{r}\otimes B_{m})\widetilde{X},I_{R}\otimes\iota_{m})$
cannot have full column rank (this follows from $(m-1)^{-1}x+(I_{r}\otimes
B_{m})x\in\operatorname{col}(I_{R}\otimes\iota_{m})$ for any $x\in
\mathbb{R}^{n}$).}

\begin{mycomment}
			if R=1 Condition \ref{assum id} violated, and $rank(\widetilde{X},W%
			\widetilde{X})<2\tilde{k}$ if $\tilde{k}>1$
		\end{mycomment}

\begin{mycomment}
			....(if R=1, col(X) is spanned by $\iota_{n}$ and something in the orth
			complement of $\iota_{n}$). ....(bramoulle sec 2.4.1.2)
		\end{mycomment}

\item[(ii)] The network fixed effects model (\ref{network model}) with each
$W_{r}$ being the symmetric or row-normalized adjacency matrix of a complete
bipartite network, and with contextual effects. In this case, the matrix $X$
is given by $(\widetilde{X},W\widetilde{X},\bigoplus_{r=1}^{R}\iota_{m_{r}})$,
and has therefore $k=R+2\tilde{k}$ columns, with $0\leq\tilde{k}<R$%
.\hfill\qed

\begin{mycomment}
			useful to rule out R=1 because in that case $rankX<k$ if $\tilde{k}>1$, and
			Condition \ref{assum id} violated
		\end{mycomment}

\begin{mycomment}
			(.....bram discuss identif after removal FE
			in this model) (see complete-bipartite-CBG-assumption-1.m).
		\end{mycomment}

\item[(iii)] A Group Interaction model with group specific regression
coefficients (denoted by $\tilde{\beta}_{r}$) and group fixed effects. The
model equation is $\boldsymbol{y}_{r}=\lambda B_{m_{r}}\boldsymbol{y}%
_{r}+\widetilde{X}_{r}\tilde{\beta}_{r}+\alpha_{r}\iota_{m_{r}}+\sigma
\boldsymbol{\varepsilon}_{r},$ for $r=1,\ldots,R,$ where the regressor matrix
$\widetilde{X}_{r}$ is $m_{r}\times k_{r}$, with $0\leq k_{r}<m_{r}$. In this
case, the matrix $X$ is given by $\bigoplus_{r=1}^{R}(\widetilde{X}_{r}%
,\iota_{m_{r}})$, and has therefore $k=R+\sum_{r=1}^{R}k_{r}$ columns.

\begin{mycomment}
	remember we need to make sure k<n (don't need to state this)
\end{mycomment}\begin{mycomment}
			R=1 Condition \ref{assum id} violated
		\end{mycomment}

\end{enumerate}
\end{example2}

\begin{example2}
\label{exa netwFE}The condition $\mathrm{rank}(X,WX)=k$ is satisfied in the
following network autoregressions with fixed effects and no regressors (i.e.,
the matrix $X$ contains only the dummies corresponding to the fixed
effects):\footnote{In case (i) Condition \ref{assum id} is satisfied for any
$W$. In cases (ii) and (iii) Condition \ref{assum id} is violated if and only
if $W$ equals the interaction matrix of a Balanced Group Interaction model
(i.e., if and only if $W=I_{T}\otimes B_{N}$ and $W=I_{R}\otimes B_{m}$ in
cases (ii) and (iii), respectively).}

\begin{enumerate}
\item[(i)] The one-way model of Example \ref{exa panel} with no regressors and
time invariant interaction matrix, as, for instance, in \cite{Robinson2015}.
In this case, letting $W_{t}=W^{\ast}$ for each $t=1,\ldots,T$, we have
$W=I_{T}\otimes W^{\ast}$, and $X$ contains only the individual fixed effects,
i.e., $X=\iota_{T}\otimes I_{N}$, so that $k=N$. Since $WX=\iota_{T}\otimes
W^{\ast}$, it follows that $\operatorname{rank}(X,WX)=\operatorname{rank}%
(\iota_{T}\otimes I_{N},\iota_{T}\otimes W^{\ast})=\operatorname{rank}%
(\iota_{T}\otimes(I_{N},W^{\ast}))=k.$

\begin{mycomment}
			here it is assumed $T>1$ but no need to say that
		\end{mycomment}

\begin{mycomment}
			......for the case of panel SAR with only individual fixed effects see individual\_fixed\_effects\_identif\_assumption\_1.m
		\end{mycomment}

\begin{mycomment}
see individual\_fixed\_effects\_identif\_assumption\_1
\end{mycomment}

\item[(ii)] The network fixed effects model of Example \ref{exa network} with
no regressors and all matrices $W_{r}$'s being row-stochastic (a matrix is
said to be row-stochastic if all its row sums are 1). In this case,
$W=\bigoplus_{r=1}^{R}W_{r}$ and $X$ contains only the network fixed effects,
i.e., $X=\bigoplus_{r=1}^{R}\iota_{m_{r}}$, so that $k=R$. Since each $W_{r}$
is row-stochastic, $\operatorname{rank}(X,WX)=\operatorname{rank}%
(\bigoplus_{r=1}^{R}\iota_{m_{r}},\bigoplus_{r=1}^{R}\iota_{m_{r}})=k.$ Note
that, when $R=1$, this reduces to an intercept-only network autoregression
(\ref{SLM}) with row-stochastic interaction matrix.

\item[(iii)] When $r$ is time, case (ii) also covers the case of a panel data
model with time fixed effects. Hence, putting cases (i) and (ii) together,
another example when $\mathrm{rank}(X,WX)=k$ is the two-way model of Example
\ref{exa panel} with no regressors (i.e., $X$ contains $k=N+T-1$ columns of
$(\iota_{T}\otimes I_{N},I_{T}\otimes\iota_{N})$) and all matrices $W_{t}$'s
being row-stochastic.\hfill\qed

\end{enumerate}
\end{example2}

Examples \ref{exa GI failure}--\ref{exa netwFE} contain several cases in which
$\mathrm{rank}(X,WX)=k$ and therefore $\lambda$ and $\beta$ cannot be
identified from $\mathrm{E}(\boldsymbol{y})$ alone. The identification
prospects in such cases are markedly different depending on whether Condition
\ref{assum id} is satisfied (Examples \ref{exa netw under cond 1} and
\ref{exa netwFE}) or not (Examples \ref{exa GI failure} and
\ref{exa CBG failure}). In the former case, identification can be achieved,
for example, by imposing higher moments restrictions (see Section
\ref{sec id II moment}). In the latter case, the identification problem is
deeper, and a solution would require more drastic changes to the model (see
Section \ref{sec inv}).

\begin{mycomment}
for nonexistence var 2SLS with 2 instruments see check\_2SLS.m
\end{mycomment}

\begin{mycomment}
	SLM\_MLE\_OLS\_2SLS\_diff\_distrib.m
\end{mycomment}

\begin{mycomment}
	WHAT HAPPENS CLOSE TO THE CASES WHERE THERE ARE PROBLEMS??? FOR EXAMPLE WHAT
	HAPPENS IF $M_{X}W$ IS CLOSE TO $0$ - see Xfixed-pert for CBG in plot-lik-score-SLM-with-adj.m
\end{mycomment}

\subsection{\label{sec id II moment}Identifiability from first two moments}

Proposition \ref{lemma identif mean} gives a condition for $\lambda$ and
$\beta$ to be generically identified when the model only specifies the first
moment of $\boldsymbol{y}$. When that condition fails, identification may be
achieved by imposing further restrictions on the model. The simplest of such
restrictions is $\mathrm{var}(\boldsymbol{\varepsilon})=I_{n}$, in which case
identification can occur via the first two moments of $\boldsymbol{y}$.

To see this, it is convenient to focus on a parameter space for $\lambda$ that
is smaller than the unrestricted parameter space $\Lambda_{\mathrm{u}%
}\coloneqq\{\lambda\in\mathbb{R}:\det(S(\lambda))\neq0\}$. Indeed, the
parameter space for $\lambda$ is usually taken to be a much smaller set than
$\Lambda_{\mathrm{u}}$. Consider the case when $W$ has at least one (real)
negative eigenvalue and at least one (real) positive eigenvalue. This is
typically satisfied in both applications and theoretical studies. Denote the
smallest real eigenvalue of $W$ by $\omega_{\min}$, and, without loss of
generality, normalize the largest real eigenvalue to 1. The parameter space
for $\lambda$ is often restricted to the largest interval containing the
origin in which $S(\lambda)$ is nonsingular, that is,
\[
\Lambda\coloneqq(\omega_{\min}^{-1},1),
\]
or a subset thereof (possibly independent of $n$) such as $(-1,1)$. Without
such parameter space restrictions, the models are believed to be too erratic
to be useful in practice, and $\lambda$ is difficult to interpret.

\begin{mycomment}
Note that the condition that $W$ has at least one negative eigenvalue and at
least one positive eigenvalue rules out the case when $W$ is a scalar multiple
of $I_{n}$, which trivially leads to non-identification of $\lambda,\beta$
from the first moment, or of $(\lambda,\beta,\sigma)$ from the first two moments.
\end{mycomment}

\begin{proposition}
\label{lemma identif var}In the network autoregression (\ref{SLM}) assume that
$\mathrm{var}(\boldsymbol{\varepsilon})=I_{n}$ and that $W$ has at least one
negative eigenvalue and at least one positive eigenvalue. The parameter
$(\lambda,\beta,\sigma)$ is identified on $\Lambda\times\mathbb{R}^{k}%
\times(0,\infty)$.
\end{proposition}

\begin{mycomment}
	seems to me for identifiability we can leave var matrix unknown. for estimation we would have to estimate it
\end{mycomment}

In Proposition \ref{lemma identif var}, the parameters $\lambda$ and $\sigma$
are identified from $\mathrm{var}(\boldsymbol{y})$. Once $\lambda\ $is
identified, $\beta$ can be identified from the first moment $\mathrm{E}%
(\boldsymbol{y})=(I_{n}-\lambda W)^{-1}X\beta$ (because $X$ has full rank). It
should be noted that the restriction $\mathrm{var}(\boldsymbol{\varepsilon
})=I_{n}$ is imposed only for simplicity, and one could certainly study
identification under more general parametric structures for $\mathrm{var}%
(\boldsymbol{\varepsilon})$. \begin{mycomment}
	in first subm I had this, but removed following one comment by AE:
is by no means necessary for
identification from $\mathrm{var}(\boldsymbol{y})$. Indeed, one could assume
some parametric structure for $\mathrm{var}(\boldsymbol{\varepsilon})$, say
$\mathrm{var}(\boldsymbol{\varepsilon})=\Sigma(\eta)$, and study
identifiability of the parameter $(\lambda,\sigma,\eta)$ from $\mathrm{var}%
(\boldsymbol{y})=\sigma^{2}(I_{n}-\lambda W)^{-1}\Sigma(\eta)(I_{n}-\lambda
W^{\prime})^{-1}$, but we refrain from doing this here.
\end{mycomment}

\begin{mycomment}
If one can identify $\mathrm{var}(\boldsymbol{\varepsilon})$ independently of $\lambda$ and
$\sigma^{2}$, then $\lambda$ and $\sigma^{2}$ can be identified by an obvious
extension of Proposition \ref{lemma identif var}, in which $I_{n}$ is replaced by
the value of $\mathrm{var}(\boldsymbol{\varepsilon})$. This is unrealistic! proof:
proof of Proposition \ref{lemma identif var} with $\mathrm{var}(\boldsymbol{\varepsilon}
)=\Sigma.$ This proof is similar to the proof of Lemma 4.2 in
\cite{Preinerstorfer2015}. Assume that $\Sigma\coloneqq\mathrm{var}%
(\boldsymbol{\varepsilon})$ is positive definite and does not depend on $\lambda$ and
$\sigma^{2}$. We establish that, if $\tilde{\sigma}^{2}S^{\prime}%
(\lambda)\Sigma^{-1}S(\lambda)=\sigma^{2}S^{\prime}(\tilde{\lambda}%
)\Sigma^{-1}S(\tilde{\lambda})$ for any two parameter values $(\lambda
,\sigma^{2}),(\tilde{\lambda},\tilde{\sigma}^{2})\in\Lambda\times(0,\infty)$,
then $(\lambda,\sigma^{2})=(\tilde{\lambda},\tilde{\sigma}^{2})$. The
maintained assumption that $W$ has at least one negative eigenvalue and at
least one positive eigenvalue guarantees the existence of a nonzero vector
$f\in\mathrm{null}(W-I_{n})$ and a nonzero vector $g\in\mathrm{null}%
(W-\omega_{\min}I_{n})$. Multiplying both sides of the equality $\tilde
{\sigma}^{2}S^{\prime}(\lambda)\Sigma^{-1}S(\lambda)=\sigma^{2}S^{\prime
}(\tilde{\lambda})\Sigma^{-1}S(\tilde{\lambda})$ by $f^{\prime}$ on the left
and $f$ on the right gives $\tilde{\sigma}^{2}\left(  1-\lambda\right)
^{2}f^{\prime}\Sigma^{-1}f=\sigma^{2}(1-\tilde{\lambda})^{2}f^{\prime}%
\Sigma^{-1}f$. Since $1-\lambda>0$ for any $\lambda\in\Lambda$ , and
$f^{\prime}\Sigma^{-1}f\neq0,$ the last equality is equivalent to
$\tilde{\sigma}/\sigma=(1-\tilde{\lambda})/\left(  1-\lambda\right)  $.
Repeating with $g$ in place of $f$ gives $\tilde{\sigma}/\sigma=(1-\tilde
{\lambda}\omega_{\min})/\left(  1-\lambda\omega_{\min}\right)  $. Thus, we
must have $\left(  1-\lambda\omega_{\min}\right)  /\left(  1-\lambda\right)
=(1-\tilde{\lambda}\omega_{\min})/(1-\tilde{\lambda})$. Since the function
$\lambda\mapsto\left(  1-\lambda\omega_{\min}\right)  /\left(  1-\lambda
\right)  $ is strictly increasing, we have $\lambda=\tilde{\lambda}$, and
hence $\sigma^{2}=\tilde{\sigma}^{2}$.
\end{mycomment}

At this point, it is worth considering the \textit{network} (or spatial)
\textit{error model}
\begin{equation}
\boldsymbol{y}=X\beta+\boldsymbol{u},\text{ }\boldsymbol{u}=\lambda
W\boldsymbol{u}+\sigma\boldsymbol{\varepsilon}, \label{SEM}%
\end{equation}
even though this specification is considerably less popular than the network
autoregression (\ref{SLM}) in economic applications. The same set of
assumptions as in the paragraph after equation (\ref{SLM}) will be maintained
for model (\ref{SEM}). Proposition \ref{lemma identif var} also applies to the
network error model, because equations (\ref{SLM}) and (\ref{SEM}) imply the
same variance structure for $\boldsymbol{y}$, and $\beta$ is trivially
identified from the first moment in this model.

On the other hand, in the network error model (\ref{SEM}) the first moment of
$\boldsymbol{y}$, $X\beta$, does not depend on $\lambda$, and hence cannot
identify $\lambda$. Proposition \ref{lemma identif mean} can be interpreted as
saying that $(\lambda,\beta)$ cannot be identified from $\mathrm{E}%
(\boldsymbol{y})$ in a network autoregression that \textquotedblleft
behaves\textquotedblright\ like a network error model. This point is made
precise by the following argument. If $\mathrm{rank}(X,WX)=k$, there exists a
unique $k\times k$ matrix $A$ such that $WX=XA$, and hence $S^{-1}%
(\lambda)X=X(I_{k}-\lambda A)^{-1},$ for any $\lambda$ such that $S(\lambda)$
is invertible (it is easily seen that the eigenvalues of $A$ are eigenvalues
of $W$, and therefore $I_{k}-\lambda A$ is invertible if $S(\lambda)$ is). It
follows that, when $\mathrm{rank}(X,WX)=k$, the network autoregression
$\boldsymbol{y}=S^{-1}(\lambda)X\beta+\sigma S^{-1}(\lambda
)\boldsymbol{\varepsilon}$ can be written as $\boldsymbol{y}=X(I_{k}-\lambda
A)^{-1}\beta+\sigma S^{-1}(\lambda)\boldsymbol{\varepsilon}$, which is a
network error model with regression coefficients $(I_{k}-\lambda A)^{-1}\beta
$.\footnote{According to Lemma \ref{lemma same prof lik} in Appendix
\ref{sec proofs}, the model $\boldsymbol{y}=X(I_{k}-\lambda A)^{-1}%
\beta+\sigma S^{-1}(\lambda)\boldsymbol{\varepsilon}$ has the same profile
quasi log-likelihood $l(\lambda,\sigma^{2})$ as model (\ref{SEM}), even
though, clearly, the MLE of $\beta$ will be different in the two models.}

\subsection{\label{sec literat}Connection with existing results}

This section discusses connections between Propositions
\ref{lemma identif mean} and \ref{lemma identif var} and some related results
available in the literature.

\begin{mycomment}
	It is useful to briefly compare Proposition \ref{lemma identif mean} with some
	related results available in the literature, obtained by two different approaches.
\end{mycomment}

\subsubsection{\label{sec bram}Identification from reduced form parameter}

In the social network literature, identification of the structural parameters
in model (\ref{SLM}) is typically established by verifying that those
parameters can be uniquely recovered from the reduced form parameters
\citep[e.g.,][]{Bramoulle2009,Blume2011,Kwok19}. Such a strategy relies on the
reduced form parameters being identified, which (in the case of a fixed $W$)
would typically require individuals being observed repeatedly, over time or
some other dimension.\footnote{\label{footnote reduced}Assuming, for
simplicity, that we have only one covariate, $x$, the reduced form of the
network autoregression $\boldsymbol{y}=\lambda W\boldsymbol{y}+\beta
x+\sigma\boldsymbol{\varepsilon}$ is $\boldsymbol{y}=\Pi x+\sigma(I-\lambda
W)^{-1}\boldsymbol{\varepsilon},$ where $\Pi\coloneqq\beta(I-\lambda W)^{-1}$
is an $n\times n$ matrix and therefore cannot, without further restrictions,
be identified from the distribution of a single realization of the vector
$\boldsymbol{y}$.} Hence, identification via reduced form parameters may not
be appropriate in applications where a single observation of a network is
available. Proposition \ref{lemma identif mean} can establish identifiability
not only when repeated observations are available (in which case $W$ is block
diagonal with identical blocks), but also when a single observation of the
network is available. The following example considers a case when parameters
can only be identified with repeated observations.

\begin{example2}
\label{exa repeated} Consider the row-normalized or symmetric Complete
Bipartite model of Example \ref{exa CBG} with an intercept, a regressor, and a
contextual effect term, so that $X=\left(  \iota_{n},x,Wx\right)  $ for some
arbitrary vector $x\in\mathbb{R}^{n}$ (such that $X$ is full rank). This model
is a particular case of equation (1) in \cite{Bramoulle2009}. Because the
matrices $I_{n}$, $W$, $W^{2}$ are linearly independent, Proposition 1 in
\cite{Bramoulle2009} establishes that the parameters $\lambda$ and $\beta$ are
identified from an i.i.d.\ sample of observations from the model, as long as
$\beta_{2}\lambda+\beta_{3}\neq0$.\footnote{\label{footnote generic}Parameter
restrictions such as $\beta_{2}\lambda+\beta_{3}\neq0$ do not appear in
Proposition \ref{lemma identif mean}, due to the fact that generic
identification is considered there.} However, in this model $\mathrm{rank}%
(X,WX)=k$, and therefore $\lambda$ and $\beta$ cannot be identified from a
single observation of the model (without assumptions beyond $\mathrm{E}%
(\boldsymbol{\varepsilon})=0$), by Proposition \ref{lemma identif mean}%
.\footnote{In fact, Condition \ref{assum id} fails in this model (see Example
\ref{exa CBG failure app} in Appendix \ref{app further} with a number of
partitions equal to two), which implies $\mathrm{rank}(X,WX)=k$.} \hfill\qed

\begin{mycomment}
\footnote{We stress that non-identification here is for any
$x$, not for generic $X$. As discussed above, Proposition \ref{lemma identif mean}
implies identification for generic $X$.}
\end{mycomment}

\end{example2}

To understand why, in Example \ref{exa repeated}, identifiability requires
repeated observations of the network it is helpful to apply Proposition
\ref{lemma identif mean} to the case in which we observe the bipartite network
$R\geq1$ times. Note that $R$ observations of a network autoregression with
$X=\left(  \iota_{n},x,Wx\right)  $ correspond to a network autoregression
with interaction matrix $W^{\ast}=I_{R}\otimes W$ and regressor matrix
$X^{\ast}=\left(  \iota_{nR},x^{\ast},W^{\ast}x^{\ast}\right)  $ for some
$x^{\ast}\in\mathbb{R}^{nR}$ (the regressor $x$ is not kept constant across
the $R$ repetitions). When $W$ is the row-normalized or symmetric Complete
Bipartite model, $\mathrm{rank}(X^{\ast},W^{\ast}X^{\ast})>k$ if and only if
$R>1$. That is, Proposition \ref{lemma identif mean} establishes that
identification is achieved if and only if $R>1$.

The applicability to the case of a single observation of a network is the most
important difference between Proposition \ref{lemma identif mean} and
Proposition 1 of \cite{Bramoulle2009}. Note also that, contrary to
\cite{Bramoulle2009}'s result, Proposition \ref{lemma identif mean} does not
restrict attention to the case when $X$ contains contextual effects; our
results can be used for that case, but also when no contextual effects are
included, or only some contextual effects are included.

\subsubsection{\label{sec asy id}Asymptotic identification}

The MLE that is typically used for a network autoregression is the one based
on the likelihood that would obtain if $\boldsymbol{\varepsilon}$ were
distributed as $\mathrm{N}(0,I_{n})$. Following common usage, we refer to this
specific quasi MLE (QMLE) simply as \textit{the} QMLE. \cite{Lee2004} studies
asymptotic properties of the QMLE. The condition $\mathrm{rank}(X,WX)>k$
appearing in Proposition \ref{lemma identif mean} can be interpreted as a
finite sample counterpart of Assumption 8 in \cite{Lee2004}. Indeed, under the
latter assumption (and other regularity assumptions) the limit of the Gaussian
quasi-likelihood has a unique maximum at the true value of the parameters,
which is a sufficient condition for identification; see \cite{NeweyMcFadden94}%
. Similarly, Conditions for $(\lambda,\sigma)$ to be identified from
$\mathrm{var}(\boldsymbol{y})$ can be seen as finite sample counterparts of
Assumption 9 in \cite{Lee2004}. \begin{mycomment}
	For the condition to be also necessary, however, correct specification of the likelihood is required [[[[in general???]]]])................
\end{mycomment}

\subsubsection{Identification from second moment}

Proposition \ref{lemma identif var} complements two results available in the
literature that are concerned with identifiability from $\mathrm{var}%
(\boldsymbol{y})$ on a parameter space for $\lambda$ different from $\Lambda$.
Firstly, Proposition \ref{lemma identif var} extends Lemma 4.2 in
\cite{Preinerstorfer2015}, which establishes identification of $(\lambda
,\sigma)$ from $\mathrm{var}(\boldsymbol{y})$ on $(0,1)\times(0,\infty)$.
Secondly, Lemma 4 in \cite{LeeYu2015} gives a sufficient condition for
$(\lambda,\sigma)$ to be identified from $\mathrm{var}(\boldsymbol{y})$ on
$\Lambda_{\mathrm{u}}\times(0,\infty)$, namely that the matrices $I_{n}$,
$W+W^{\prime}$ and $W^{\prime}W$ are linearly independent. It is instructive
to look at an example in which $(\lambda,\sigma)$ is identified from
$\mathrm{var}(\boldsymbol{y})$ on $\Lambda_{\mathrm{u}}\times(0,\infty)$ but
not on $\Lambda_{\mathrm{u}}\times(0,\infty)$, and therefore identification
can be established by Proposition \ref{lemma identif var} but not by Lemma 4
in \cite{LeeYu2015}.

\begin{example2}
\label{Exa BGI var}Consider a balanced group interaction model (see Example
\ref{exa GI}) with $\mathrm{var}(\boldsymbol{\varepsilon})=I_{n}$. By
Proposition \ref{lemma identif var}, $(\lambda,\sigma)$ is identified on
$\Lambda\times(0,\infty)$ (and hence on any subset thereof). On the other
hand, Lemma 4 in \cite{LeeYu2015} cannot establish identifiability, or
non-identifiability, on $\Lambda\times(0,\infty)$ because the matrices $I_{n}%
$, $W+W^{\prime}$ and $W^{\prime}W$ are not linearly independent when
$W=I_{R}\otimes B_{m}$. Indeed, when $W=I_{R}\otimes B_{m}$, $\sigma_{1}%
^{2}\left(  S^{\prime}(\lambda_{1})S(\lambda_{1})\right)  ^{-1}=\sigma_{2}%
^{2}\left(  S^{\prime}(\lambda_{2})S(\lambda_{2})\right)  ^{-1}$ if and only
if $\sigma_{2}^{2}=m^{2}\sigma_{1}^{2}/(2\lambda_{1}+m-2)^{2}$ and
$\lambda_{2}=-((m-2)\lambda_{1}+2(1-m))/(2\lambda_{1}+m-2)$, which shows that
$(\lambda,\sigma)$ is not identifiable on $\Lambda_{\mathrm{u}}\times
(0,\infty)$. Note however that $\lambda_{2}\notin\Lambda$ if $\lambda_{1}%
\in\Lambda$, which confirms that $(\lambda,\sigma)$ is identifiable on the
smaller set $\Lambda\times(0,\infty)$.\hfill\qed

\end{example2}

Exploiting second moment restrictions to achieve identifiability has been
considered in many areas of econometrics
\citep[see, e.g.,][in the context of DSGE models]{KomunjerNg11}, and in
particular in the peer effects literature; see, for instance,
\cite{Graham2008}, Theorem 3.2 in \cite{Davezies09}, \cite{Rose2017}, and
\cite{Liu2017}. The latter paper studies identifiability in the context of the
model in Example \ref{exa netw under cond 1}(i) above.

\begin{mycomment}
	(note that the model in the footnote  in the example above is
	different from the one in case (b.iii) in Section \ref{sec id first} because
	it has the intercept $\iota_{nR}$ instead of the group intercepts
	$I_{R}\otimes\iota_{n}$)
\end{mycomment}

\begin{mycomment}
in the prev example ...allowing $X$ to vary across the $R$ observations
\end{mycomment}

\begin{mycomment}	
	in the example 2SLS defined only with repeated obs
\end{mycomment}

\begin{mycomment}	
	- so, are we saying that with $W^{\ast}=I_{R}\otimes W$ and $X^{\ast}=\left(
	\iota_{nR},x,Wx\right)  $ we can always identify from first moment if $I_{n}$,
	$W$, $W^{2}$ are l.i. and $R>1$? yeah I\ think that's true: $\mathrm{rank}%
	(X^{\ast},W^{\ast}X^{\ast})=\mathrm{rank}\left(  \iota_{nR},x,W^{\ast
	}x,W^{\ast}\iota_{nR},W^{\ast}x,W^{\ast2}x\right)  =[$with $W$
	row-stoch$]\mathrm{rank}\left(  \iota_{nR},x,W^{\ast}x,W^{\ast2}x\right)  $
	- A1 violated so MLE constant (=0) and OLS degenerate.
	- concerning the model with $R$ observations, if $\iota_{nR}$ is replaced by
	the FE $I_{R}\otimes\iota_{n}$ then $\mathrm{rank}(X^{\ast},W^{\ast}X^{\ast
	})=k$ for any $R$ so no identif from first moment even under repeated
	observations (this example discussed above, where we also say that Assumption
	1 violated only if R=1)). For this model, non identification (after removal of
	FE)\ according to Bram, because $I,W^{\ast},W^{\ast2},W^{\ast3}$ are lin dep
	(in this case $\mathrm{rank}(X,WX)>k$)
\end{mycomment}

\begin{mycomment}
	FOR\ MYSELF: The presence of two distinct eigenvalues of $W$ is not sufficient
	for identifiability over the set of $\lambda\ $such that $S(\lambda)$ is
	nonsingular, because $\left\vert \left(  1-\lambda_{2}\right)  /\left(
	1-\lambda_{1}\right)  \right\vert =\left\vert \left(  1-\lambda_{2}%
	\omega_{\min}\right)  /\left(  1-\lambda_{1}\omega_{\min}\right)  \right\vert
	$ admits the sol $\lambda_{1}=\frac{\lambda_{2}\omega+\lambda_{2}-1}%
	{2\lambda_{2}\omega-\omega}-1$
	i.e. with 3 distinct eigenv we have
	If $\sigma_{1}^{2}(S^{\prime}(\lambda_{1})S_{\lambda_{1}})^{-1}=\sigma
	_{2}^{2}(S^{\prime}(\lambda_{2})S_{\lambda_{2}})^{-1}$ holds for $\lambda
	_{i}\ $such that $S_{\lambda_{i}}$ is nonsingular and $0<\sigma_{i}<\infty$
	($i=1,2$), then $\lambda_{1}=\lambda_{2}$ and $\sigma_{1}=\sigma_{2}.\bigskip$
	$\sigma_{2}^{2}/\sigma_{1}^{2}=\left(  1-\lambda_{2}\right)  ^{2}/\left(
	1-\lambda_{1}\right)  ^{2}$
	$\sigma_{2}^{2}/\sigma_{1}^{2}=\left(  1-\lambda_{2}\omega_{\min}\right)
	/\left(  1-\lambda_{1}\omega_{\min}\right)  $
	$\sigma_{2}^{2}/\sigma_{1}^{2}=\left(  1-\lambda_{2}\omega\right)  /\left(
	1-\lambda_{1}\omega\right)  $
	\bigskip
	1) $\left(  1-\lambda_{2}\right)  /\left(  1-\lambda_{1}\right)  =\pm\left(
	1-\lambda_{2}\omega_{\min}\right)  /\left(  1-\lambda_{1}\omega_{\min}\right)
	$
	2) $\left(  1-\lambda_{2}\right)  /\left(  1-\lambda_{1}\right)  =\pm\left(
	1-\lambda_{2}\omega\right)  /\left(  1-\lambda_{1}\omega\right)  $
	1) with + is impossible, but is possible with -, in which case we have
	$\lambda_{1}=\frac{\lambda_{2}\omega+\lambda_{2}-1}{2\lambda_{2}\omega-\omega
	}-1$ (see identification\_3\_eigenv.mw). We cannot have $\lambda_{1}%
	=\frac{\lambda_{2}\omega+\lambda_{2}-1}{2\lambda_{2}\omega-\omega}-1$ and
	$\lambda_{1}=\frac{\lambda_{2}\omega_{\min}+\lambda_{2}-1}{2\lambda_{2}%
		\omega_{\min}-\omega_{\min}}-1$ for $\omega\neq\omega_{\min}$
\end{mycomment}

\begin{mycomment}
	$%
	\begin{array}
	[c]{cc}%
	1 & -1\\
	0 & 1
	\end{array}
	$, eigenvalues: $1$\bigskip
	$\frac{d}{d\rho}\frac{1+\rho^{2}b}{\left(  1-\rho\omega\right)  ^{2}%
	}=\allowbreak-\frac{2}{\left(  \omega\rho-1\right)  ^{3}}\left(  \omega
	+b\rho\right)  $
	$\frac{d}{\diff\lambda}\frac{1-\lambda\omega_{2}}{1-\lambda\omega_{1}}%
	=\allowbreak\frac{1}{\left(  \lambda\omega_{1}-1\right)  ^{2}}\left(
	\omega_{1}-\omega_{2}\right)  $
	\bigskip$1-\lambda\omega_{1}>0$, Solution is: $\left\{
	\begin{array}
	[c]{ccc}%
	\left(  -\infty,\frac{1}{\omega_{1}}\right)  & \text{if} & 0<\omega_{1}%
	\wedge\frac{1}{\omega_{1}}\in%
	\mathbb{R}%
	\\
	\left(  \frac{1}{\omega_{1}},\infty\right)  & \text{if} & \omega_{1}%
	<0\wedge\frac{1}{\omega_{1}}\in%
	\mathbb{R}%
	\\
	\emptyset & \text{if} & \frac{1}{\omega_{1}}\in%
	\mathbb{R}
	\wedge\lnot\omega_{1}\in%
	\mathbb{R}%
	\\%
	\mathbb{R}%
	& \text{if} & \omega_{1}=0\\
	\left\{  0\right\}  & \text{if} & \omega_{1}\in%
	\mathbb{C}
	\setminus%
	\mathbb{R}
	\wedge\frac{1}{\omega_{1}}\in%
	\mathbb{C}
	\setminus%
	\mathbb{R}%
	\end{array}
	\right.  \allowbreak$
\end{mycomment}

\section{\label{sec inv}Invariance}

So far, we have considered identifiability of a parameter $\theta$ from the
distribution $P_{\theta}$ of an observable random vector $\boldsymbol{y}$.
Sometimes, it may be appropriate to consider identifiability from some
transformation of $\boldsymbol{y}$. Such a transformation might be dictated by
the desire to eliminate some nuisance parameters, or, more generally, by
invariance considerations \citep[see, e.g.,][]{ChamberlainMoreira2009}.
Suppose, for example, that $\theta$ is partitioned as $(\theta_{1}^{\prime
},\theta_{2}^{\prime})^{\prime}$, where $\theta_{2}$ is not of direct
interest. Particularly when the dimension of $\theta_{2}$ is large compared to
the sample size, one may want to consider identification from a transformation
of $\boldsymbol{y}$ whose distribution does not depend on $\theta_{2}$. In
general, if $\theta_{1}$ is identified from the distribution of
$\boldsymbol{y}$ it is also identified from the distribution of the
transformation of $\boldsymbol{y}$. However, as we shall see in this section,
this is not always the case. To analyze this point, we will need to discuss
the full identifiability content of Condition \ref{assum id}. We have seen in
Section \ref{sec id first} that, in a network autoregression, Condition
\ref{assum id} is not necessary for identifiability from the distribution of
$\boldsymbol{y}$; more precisely, it is necessary for identifiability from the
first moment of $\boldsymbol{y}$, but not for identifiability from the second
moment. Nonetheless, we will show that when Condition \ref{assum id} fails it
is impossible to conduct inference that respects the symmetry properties of
the model. More precisely, when Condition \ref{assum id} fails the model is
invariant under certain transformations of the sample space, but
identifiability is impossible from any function of the data that is invariant
under those transformations. This result requires some general group
invariance notions \citep[e.g.,][Chapter 6]{Lehmann2005}, which are reviewed
in Section \ref{sec notions inv}, and then applied to network autoregressions
in Section \ref{sec inv SAR}. Section \ref{sec role} contains the main
identifiability result, and Section \ref{sec lik} discusses some implications
for likelihood inference.

\subsection{\label{sec notions inv}General invariance notions}

Let $\mathcal{G}$ be a group of one-to-one functions (transformations) from a
space $\mathcal{S}$ into itself, with the group operation being the
composition of functions. The group $\mathcal{G}$ induces a partition of
$\mathcal{S}$ into equivalence classes, called orbits, with two elements of
$\mathcal{S}$ being in the same orbit if there exists an element of
$\mathcal{G}$ transforming one element into the other. The orbit of an element
$x\in\mathcal{S}$ is therefore the set $\left\{  g(x):g\in\mathcal{G}\right\}
$. A function on $\mathcal{S}$ is said to be invariant under $\mathcal{G}$ (or
$\mathcal{G}$-invariant) if it is constant on the orbits of $\mathcal{G}$. A
function on $\mathcal{S}$ is said to be a maximal invariant under
$\mathcal{G}$ if it is invariant and takes different values on each orbit. A
necessary and sufficient condition for a function on $\mathcal{S}$ to be
invariant under $\mathcal{G}$ is that it depends on $x\in\mathcal{S}$ only
through a maximal invariant under $\mathcal{G}$.

\begin{example2}
\label{exa scale}(Scale invariance) Consider the group $\mathcal{G}%
=\{g_{\kappa}:\kappa>0\}$, where $g_{\kappa}$ is the function $y\mapsto\kappa
y$ from $\mathbb{R}^{n}$ to $\mathbb{R}^{n}$. A maximal invariant under
$\mathcal{G}$ is $y/\left\Vert y\right\Vert $, where $\left\Vert y\right\Vert
\coloneqq\sqrt{y^{\prime}y}$, and hence a function on $\mathbb{R}^{n}$ is
$\mathcal{G}$-invariant is and only if it depends on $y$ only through
$y/\left\Vert y\right\Vert $.\footnote{Here we use the convention that
$y/\left\Vert y\right\Vert =0$ if $y=0$. Invariance of $y/\left\Vert
y\right\Vert $ holds because $y/\left\Vert y\right\Vert =\kappa y/\left\Vert
\kappa y\right\Vert $ for any $\kappa>0$, maximality because, for any
$y,\tilde{y}\in\mathbb{R}^{n}$, $y/\left\Vert y\right\Vert =\tilde
{y}/\left\Vert \tilde{y}\right\Vert $ if and only if $\tilde{y}=ay$ for some
$a>0$.}\hfill\qed

\end{example2}

The notion of group invariance can also be applied to a statistical model,
defined to be a family of distributions on a certain sample space. In this
case the set $\mathcal{S}$ upon which the group $\mathcal{G}$ acts is the
sample space, and the functions $g$ in $\mathcal{G}$ are required to be
measurable (with respect to the sample space $\sigma$-algebra), so that when
$\boldsymbol{z}$ is a random variable with values in $\mathcal{S}$,
$g(\boldsymbol{z})$ is too. The family of distributions $\left\{  P_{\theta
}:\theta\in\Theta\right\}  $ is said to be invariant under $\mathcal{G}$ if it
is closed under the action of $\mathcal{G}$, that is, if every pair
$g,\theta\in\mathcal{G}\times\Theta$ determines a unique element in $\Theta$
denoted by $\bar{g}(\theta)$, such that when $\boldsymbol{z}$ has distribution
$P_{\theta}$, then $g(\boldsymbol{z})$ has distribution $P_{\bar{g}(\theta)}$.
Note that the definition of invariance of a family of distributions requires
$\theta$ to be identified. The functions $\theta\mapsto\bar{g}(\theta)$ form a
group acting on the parameter space, denoted by $\mathcal{\bar{G}}$.

When a model for $\boldsymbol{y}$ is invariant under a group $\mathcal{G}$, it
is natural to require that any inferential conclusion should be the same
whether $y$ or $g(y)$ is observed, for any $g\in\mathcal{G}$. Accordingly, the
data $y$ can be reduced to any $\mathcal{G}$-invariant function of $y$, with
maximal reduction being achieved by reduction to a maximal invariant. This is
what the so called principle of invariance advocates. For example, one would
typically want to restrict attention to invariant loss functions (and,
correspondingly, equivariant estimators) and, when the both the null and the
alternative hypotheses are preserved under the group $\mathcal{G}$, to
invariant tests. One fundamental results in the theory of invariance says that
the distribution of any invariant statistic depends on $\theta$ only through a
maximal invariant under $\mathcal{\bar{G}}$. Thus, in addition to a reduction
in the sample space (from the dimension of $y$ to the dimension of the maximal
invariant under $\mathcal{G}$), the principle of invariance generally also
implies a reduction in the parameter space (from the dimension of $\theta$ to
that of the maximal invariant under $\mathcal{\bar{G}}$).

\begin{example2}
\label{exa scale model}(A simple scale invariant model)\ Consider the
statistical model defined by $\boldsymbol{y}=\sigma\boldsymbol{\varepsilon}$,
where the distribution of $\boldsymbol{\varepsilon}$ depends on a parameter
$\eta$, and does not depend on the parameter $\sigma>0$. Provided that $\eta$
is identified, the model is invariant under the group of scale transformations
$\mathcal{G}=\{g_{\kappa}:\kappa>0\}$ (Example \ref{exa scale}), because if
$\boldsymbol{y}$ has distribution $P_{\sigma,\eta}$ $g(\boldsymbol{y})$ has
distribution $P_{\bar{g}(\sigma,\eta)},$ with $\bar{g}(\theta)=(\kappa
\sigma,\eta)$, for any $g\in\mathcal{G}$. The maximal invariant under
$\mathcal{\bar{G}}$ is $\eta$, and indeed the distribution of the maximal
invariant under $\mathcal{G}$, $\boldsymbol{y}/\left\Vert \boldsymbol{y}%
\right\Vert $, depends only on $\eta$.\hfill\qed

\end{example2}

\subsection{\label{sec inv SAR}Invariance of a network autoregression}

We start our discussion of the invariance properties of a network
autoregression by providing an invariance interpretation of Proposition
\ref{lemma identif mean}. For this we need to introduce a group that is often
used for regression models \citep[e.g.,][]{kariya1980}, and we need a
definition of invariance of an expectation. For a given $n\times m$ full
column rank matrix $Z$, define the group $\mathcal{G}_{Z}\coloneqq\{g_{\kappa
,\delta}:\kappa>0,\delta\in\mathbb{R}^{m}\}$, where $g_{\kappa,\delta}$
denotes the function $y\mapsto\kappa y+Z\delta$ (a one-to-one transformation
of $\mathbb{R}^{n}$), and its subgroup $\mathcal{G}_{Z}^{1}%
\coloneqq\{g_{1,\delta}:\delta\in\mathbb{R}^{m}\}$. A maximal invariant under
$\mathcal{G}_{Z}^{1}$ is $C_{Z}\boldsymbol{y}$, where $C_{Z}$ is an $\left(
n-m\right)  \times n$ matrix such that $C_{Z}C_{Z}^{\prime}=I_{n-m}$ and
$C_{Z}^{\prime}C_{Z}=M_{Z}$, and a maximal invariant under $\mathcal{G}_{Z}$
is $v(\boldsymbol{y})\coloneqq C_{Z}\boldsymbol{y}/\left\Vert C_{Z}%
\boldsymbol{y}\right\Vert $ (with the convention that $v=0$ if $C_{Z}%
y=0$).\footnote{\label{footnote max inv}This can be established by direct
verification of the definition of a maximal invariant. We provide the argument
for $\mathcal{G}_{Z}$ (the argument for $\mathcal{G}_{Z}^{1}$ is similar). The
statistic $v(y)$ is invariant because $v(\kappa y+Z\delta)=v(y)$ for any
$\kappa>0$ and any $\delta\in\mathbb{R}^{m}$, since $C_{Z}Z=0$. It takes on
different values on different orbits because, for any $y,\tilde{y}%
\in\mathbb{R}^{n}$, $v(y)=v(\tilde{y})$ if and only if $C_{Z}\tilde{y}%
=aC_{Z}y$ for some $a>0$, that is, $C_{Z}(\tilde{y}-ay)=0$, which is
equivalent to $\tilde{y}=ay+Zb$ for some $b\in\mathbb{R}^{m}$.} We say that
the expectation $\mathrm{E}_{\theta}(\boldsymbol{y})$ of a family of
distributions $\left\{  P_{\theta}:\theta\in\Theta\right\}  $ is $\mathcal{G}%
$-invariant if every pair $g,\theta\in\mathcal{G}\times\Theta$ determines a
unique $\bar{g}(\theta)$ such that $\mathrm{E}_{\theta}(g(\boldsymbol{y}%
))=\mathrm{E}_{\bar{g}(\theta)}(\boldsymbol{y})$.\footnote{Note that
$\mathcal{G}$-invariance of $\mathrm{E}_{\theta}(\boldsymbol{y})$ is necessary
but not sufficient for $\mathcal{G}$-invariance of $\left\{  P_{\theta}%
:\theta\in\Theta\right\}  $ (the latter requires that every pair $g,\theta
\in\mathcal{G}\times\Theta$ determines a unique $\bar{g}(\theta)$ such that
$\mathrm{E}_{\theta}(\varphi(g(\boldsymbol{y})))=\mathrm{E}_{\bar{g}(\theta
)}(\varphi(\boldsymbol{y}))$, for every measurable function $\varphi$).} Armed
with these definitions, the non-identifiability result in Proposition
\ref{lemma identif mean}(ii) can be seen as a consequence of the fact that,
when $\mathrm{rank}(X,WX)=k$, the expectation $\mathrm{E}_{\lambda,\beta
}(\boldsymbol{y})\coloneqq S^{-1}(\lambda)X\beta$ is $\mathcal{G}_{X}^{1}%
$-invariant.\footnote{If $\mathrm{rank}(X,WX)=k$, there exists a unique
$k\times k$ matrix $A$ such that $WX=XA$, and hence $S^{-1}(\lambda
)X=X(I_{k}-\lambda A)^{-1},$ for any $\lambda$ such that $S(\lambda)$ is
invertible (note that $I_{k}-\lambda A$ is invertible if $S(\lambda)$ is,
because the eigenvalues of $A$ must be eigenvalues of $W$). Hence
$\mathrm{E}_{\lambda,\beta}(g_{1,\delta}(\boldsymbol{y}))=\mathrm{E}_{\bar
{g}\left(  \lambda,\beta\right)  }(\boldsymbol{y})$, with $\bar{g}\left(
\lambda,\beta\right)  =(I_{k}-\lambda A)^{-1}\beta+\delta.$} This type of
invariance implies that, when $\mathrm{rank}(X,WX)=k$, the expectation
$\mathrm{E}_{\lambda,\beta}(\boldsymbol{y})$ can only identify a
$k$-dimensional parameter, not the $(k+1)$-dimensional parameter
$(\lambda,\beta)$. We will now show that, under the same rank condition, the
full family of distributions underlying a network autoregression (not the mean
only) is invariant under $\mathcal{G}_{X}^{1}$, in fact under $\mathcal{G}%
_{X}$. But before stating the result for the network autoregression, it is
helpful to consider the network error model \eqref{SEM}.

\begin{assumption}
\label{assum distrib}The distribution of $\boldsymbol{\varepsilon}$ does not
depend on the parameters $\lambda$, $\beta$, and $\sigma$.
\end{assumption}

Let $P_{\theta}$ denote the distribution for $\boldsymbol{y}$ in the network
error model, with $\theta\coloneqq(\lambda,\beta,\sigma,\eta)$, $\eta$ being a
parameter indexing the distribution of $\boldsymbol{\varepsilon}$. Under
Assumption \ref{assum distrib} and provided that $\theta$ is identified, the
network error model is $\mathcal{G}_{X}$-invariant, because $g(\boldsymbol{y}%
)$ has distribution $P_{\bar{g}(\theta)},$ with $\bar{g}(\theta)=(\lambda
,\kappa\beta+\delta,\kappa^{2}\sigma^{2},\eta)$, for any $g\in\mathcal{G}_{X}%
$. Using the same parametrization, the result for network autoregression is as follows.

\begin{mycomment}
	the following was a footnote but actually not needed as I say it earlier that ran(X,WX) is trivially satisfied when X=0.
	Note that Lemma \ref{lemma model inv} also holds for a pure
	network autoregression, by setting $X=0$.
\end{mycomment}

\begin{mycomment}
	when the model is semiparametric, check if
	I\ need to assume anything more
\end{mycomment}

\begin{mycomment}
Lemma \ref{lemma model inv} also holds for a pure network autoregression, in which case
$\operatorname{col}(X)$ is the trivial invariant subspace $\{0\}$ and
$\mathcal{G}_{X}$ is the group $\mathcal{G}_{0}$ of scale
transformations.\bigskip
\end{mycomment}

\begin{mycomment}
	\emph{see SEM\_SLM\_estimates.m}
\end{mycomment}

\begin{lemma}
\label{lemma model inv}Suppose that, in the network autoregression (\ref{SLM})
Assumption \ref{assum distrib} holds and $\theta$ is identified. The model is
$\mathcal{G}_{X}$-invariant if and only if $\mathrm{rank}(X,WX)=k$.
\end{lemma}

Of course, invariance under a certain group implies invariance under a
subgroup of that group. A subgroup of $\mathcal{G}_{X}$ that will play an
important role in Section \ref{sec role} is $\mathcal{G}_{X^{\ast}}^{1}$, for
some $n\times k^{\ast}$ submatrix $X^{\ast}$ of $X$ ($k^{\ast}\leq k$). This
is the group of transformations $y\mapsto y+X^{\ast}\delta$. For a network
autoregression, it is clear from the proof of Lemma \ref{lemma model inv} that
a sufficient condition for the model to be $\mathcal{G}_{X^{\ast}}^{1}%
$-invariant (and also $\mathcal{G}_{X^{\ast}}$-invariant) is that
$\mathrm{rank}(X^{\ast},WX^{\ast})=k^{\ast}$. Recall now that the principle of
invariance says that if a model is invariant under a group $\mathcal{G}$ then
the data should be reduced to $\mathcal{G}$-invariant functions of the data,
i.e., to functions of the data that depend on $\boldsymbol{y}$ only through
the maximal invariant under $\mathcal{G}$. In particular, imposition of
invariance under $\mathcal{G}_{X^{\ast}}^{1}$ implies that $X^{\ast}$ is
removed from the model, because the maximal invariant under $\mathcal{G}%
_{X^{\ast}}^{1}$ is $C_{X^{\ast}}\boldsymbol{y}$ and $C_{X^{\ast}}X^{\ast}=0$.
In the following example, $X^{\ast}$ is a matrix of fixed effects.

\begin{mycomment}
$\boldsymbol{y}=S^{-1}(\lambda)X^{\ast}\beta^{\ast}+S^{-1}(\lambda)X^{\ast
\ast}\beta^{\ast\ast}+\sigma S^{-1}(\lambda)\boldsymbol{\varepsilon}$
$g(\boldsymbol{y})=\kappa\boldsymbol{y+}X^{\ast}\delta$
$g(\boldsymbol{y})=\kappa S^{-1}(\lambda)X^{\ast}\beta^{\ast}+X^{\ast}%
\delta+\kappa S^{-1}(\lambda)X^{\ast\ast}\beta^{\ast\ast}+\kappa\sigma
S^{-1}(\lambda)\boldsymbol{\varepsilon}$
inv if $\operatorname{col}(S^{-1}(\lambda)X^{\ast})=\operatorname{col}%
(X^{\ast})$ iff $\operatorname{col}(S(\lambda)X^{\ast})=\operatorname{col}%
(X^{\ast})$ iff $\operatorname{col}(WX^{\ast})\subseteq\operatorname{col}%
(X^{\ast})$ iff $\mathrm{rank}(X^{\ast},WX^{\ast})=k^{\ast}$
inv if $\operatorname{col}(S^{-1}(\lambda)X^{\ast\ast})=\operatorname{col}%
(X^{\ast})$ iff $\operatorname{col}(S(\lambda)X^{\ast})=\operatorname{col}%
(X^{\ast\ast})$ iff
$\boldsymbol{y}=S^{-1}(\lambda)X_{1}\beta_{1}+S^{-1}(\lambda)X_{2}\beta
_{2}+\sigma S^{-1}(\lambda)\boldsymbol{\varepsilon}$ invariant under $y\mapsto
y+X_{1}\delta$ if (not iff prpbably?????? because I\ could also have
$\operatorname{col}(S^{-1}(\lambda)X_{2})=\operatorname{col}(X_{1})$)
$\operatorname{col}(S^{-1}(\lambda)X_{1})=\operatorname{col}(X_{1})$, or,
which is the same, $\operatorname{col}(S(\lambda)X_{1})=\operatorname{col}%
(X_{1}).$ equivalent to $\mathrm{rank}(X_{1},WX_{1})=k_{1}$. $\mathrm{rank}%
(X^{\ast},WX^{\ast})=k^{\ast}$ is not necessary for $\mathcal{G}_{X^{\ast}}%
$-invariance when $k^{\ast}\leq k$ so for example $\boldsymbol{y}=\lambda
W\boldsymbol{y}+X\beta+\sigma\boldsymbol{\varepsilon}$ is inv to $y\mapsto
y+\iota\delta$ if W is row-stoch and X includes intercept
\end{mycomment}

\begin{example2}
\label{network FE}Consider the network fixed effects model of Example
\ref{exa network}, and let $X_{\mathrm{FE}}:=\bigoplus_{r=1}^{R}\iota_{m_{r}}%
$, the $n\times R$ matrix of network fixed effects. Under Assumption
\ref{assum distrib}, and as long as each $W_{r}$ is row-stochastic, the model
is $\mathcal{G}_{X_{\mathrm{FE}}}^{1}$-invariant, because $W_{r}\iota_{m_{r}%
}=\iota_{m_{r}}$ and therefore $\mathrm{rank}(X_{\mathrm{FE}},WX_{\mathrm{FE}%
})=R$. In this case, the principle of invariance suggests that data should be
reduced to $\mathcal{G}_{X_{\mathrm{FE}}}^{1}$-invariant functions of $y$,
that is, to functions that depend on $y$ only through the maximal invariant
$C_{X_{\mathrm{FE}}}\boldsymbol{y}$ under $\mathcal{G}_{X_{\mathrm{FE}}}$.
Since $C_{X_{\mathrm{FE}}}X_{\mathrm{FE}}=0$, reduction to $\mathcal{G}%
_{X_{\mathrm{FE}}}^{1}$-invariant statistics removes the fixed effects. We
note that transformation by $C_{X_{\mathrm{FE}}}$ is equivalent to the
transformation proposed in \cite{LeeLiuLin2010} to eliminate the network fixed
effects (see Appendix \ref{app lik}). Other transformations used to remove
fixed effects in this model may or may not satisfy the principle of
invariance. For example, the transformation referred to as global differences
in \cite{Bramoulle2009} does, whereas that referred to as local differences in
the same paper does not.
\end{example2}

\subsection{\label{sec role}The role of Condition \ref{assum id}}

We are now in a position to discuss the implications of Condition
\ref{assum id}. Recall from Section \ref{sec id first} that $\mathrm{rank}%
(X,WX)=k$ if Condition \ref{assum id} is violated. Thus, by Lemma
\ref{lemma model inv} and the principle of invariance, any failure of
Condition \ref{assum id} is a case in which one would want to reduce the data
to $\mathcal{G}_{X}$-invariant functions of the data. However, the imposition
of $\mathcal{G}_{X}$-invariance causes an identifiability issue when Condition
\ref{assum id} fails. To see this, observe that if Condition \ref{assum id}
fails then $C_{X}S(\lambda)=(1-\lambda\omega)C_{X}$, and therefore
premultiplying both sides of the network autoregression equation
$S(\lambda)\boldsymbol{y}=X\beta+\sigma\boldsymbol{\varepsilon}$ by $C_{X}$
yields
\begin{equation}
C_{X}\boldsymbol{y}=\frac{\sigma}{1-\lambda\omega}C_{X}\boldsymbol{\varepsilon
}. \label{Cy eq}%
\end{equation}
Note that $\lambda$ and $\sigma$ appear together in the scale factor in front
of $C_{X}\boldsymbol{\varepsilon}$. Thus, when Assumption \ref{assum distrib}
is satisfied but Condition \ref{assum id} is not, $(\lambda,\beta,\sigma)$
cannot be separately identified from the distribution of $C_{X}\boldsymbol{y}$
and hence, since $C_{X}\boldsymbol{y}$ is a maximal invariant under
$\mathcal{G}_{X}^{1}$, cannot be identified from the distribution of any
$\mathcal{G}_{X}^{1}$-invariant statistic. Exactly the same conclusion obtains
starting from the network error model $\boldsymbol{y}=X\beta+\sigma
S^{-1}(\lambda)\boldsymbol{\varepsilon}$. The result is particularly perverse
for the network autoregression: when Condition \ref{assum id} fails, and under
Assumption \ref{assum distrib}, the model is $\mathcal{G}_{X}^{1}$-invariant
provided that its parameter $\theta$ is identifiable from the distribution of
$\boldsymbol{y}$, and yet $\theta$ cannot be identified from any
$\mathcal{G}_{X}^{1}$-invariant statistic.

It is possible to be more precise about the cause of this identification
failure. Suppose Condition \ref{assum id} is violated for some eigenvalue
$\omega$ of $W$, and let $\mathcal{\gamma}_{\omega}$ be the geometric
multiplicity of $\omega$.\footnote{Note that, for fixed $W$ and $X$, the
condition $M_{X}(\omega I_{n}-W)=0$ that leads to a violation of Condition
\ref{assum id} can be satisfied at most by one eigenvalue $\omega$. This is
because $M_{X}(\omega_{1}I_{n}-W)=M_{X}(\omega_{2}I_{n}-W)$ implies
$\omega_{1}=\omega_{2}$. Also, note that $M_{X}(\omega I_{n}-W)=0$ implies
that $\omega$ is real.} Recall from Section \ref{sec id first} that a pair
$(X,W)$ causes Condition \ref{assum id} to fail if and only if some of the
columns of $X$ span the subspace $\operatorname{col}(\omega I_{n}-W)$. Observe
that this requires $k\geq n-\mathcal{\gamma}_{\omega}$, because the dimension
of $\operatorname{col}(\omega I_{n}-W)$ is $\mathrm{rank}(\omega
I_{n}-W)=n-\mathrm{nullity}(\omega I_{n}-W)=n-\mathcal{\gamma}_{\omega}$. Let
$X_{\omega}$ be the $n\times(n-\mathcal{\gamma}_{\omega})$ matrix containing
the columns of $X$ that span $\operatorname{col}(\omega I_{n}-W)$, and reorder
the columns of $X$ as in $X=(X_{\omega},X^{\ast})$, where $X^{\ast}$ is
$n\times(k-(n-\mathcal{\gamma}_{\omega}))$, with $k-(n-\mathcal{\gamma
}_{\omega})\geq0$. Generalizing the argument leading to equation
(\ref{Cy eq}), if Condition \ref{assum id} fails then $C_{X_{\omega}}%
S(\lambda)=(1-\lambda\omega)C_{X_{\omega}}$, and therefore
\begin{equation}
C_{X_{\omega}}\boldsymbol{y}=\frac{1}{1-\lambda\omega}C_{X_{\omega}}X^{\ast
}\beta^{\ast}+\frac{\sigma}{1-\lambda\omega}C_{X_{\omega}}%
\boldsymbol{\varepsilon}, \label{Comegay}%
\end{equation}
where $\beta^{\ast}$ is the component of $\beta$ corresponding to $X^{\ast}$.
This shows that, under Assumption \ref{assum distrib}, $(\lambda,\beta
,\sigma)$ cannot be identified from the distribution of $C_{X_{\omega}%
}\boldsymbol{y}$ if Condition \ref{assum id} fails. That is, what really
causes non-identification when Condition \ref{assum id} fails is the
imposition of invariance with respect to the subgroup $\mathcal{G}_{X_{\omega
}}^{1}$ of $\mathcal{G}_{X}$, and what we said above about $\mathcal{G}%
_{X}^{1}$-invariant statistics applies to the (larger) set of $\mathcal{G}%
_{X_{\omega}}^{1}$-invariant statistics. We formally state this result in the
following theorem, and then provide an example.

\begin{theo}
\label{theo non-identif} Suppose that, in the network autoregression
(\ref{SLM}) or in the network error model (\ref{SEM}), Assumption
\ref{assum distrib} is satisfied. If Condition \ref{assum id} fails for some
eigenvalue $\omega$ of $W$, then $(\lambda,\beta,\sigma)$ cannot be identified
from the distribution of any $\mathcal{G}_{X_{\omega}}^{1}$-invariant statistic.
\end{theo}

Theorem \ref{theo non-identif} says that, for any $W$, there are matrices of
regressors that make invariant inference impossible---these are the matrices
leading to a violation of Condition \ref{assum id}, that is, the matrices
whose column space contains a subspace $\operatorname{col}(\omega I_{n}-W)$,
for some eigenvalue $\omega$ of $W$. It is worth emphasizing that this result
does not require any distributional assumption other than Assumption
\ref{assum distrib}. We provide an illustration of Theorem
\ref{theo non-identif} by revisiting a well-known identification failure in
the context of the balanced group interaction model (see Example \ref{exa GI}).

\begin{mycomment}
	The result about
	non-identifiability after reduction by $\mathcal{G}_{X}^{1}$-invariance is a
	direct consequence of the fact that the distribution of $Cy$ depends on
	$\lambda$ and $\sigma^{2}$ only through $\sigma/(1-\lambda\omega)$ (if the
	distribution of $\boldsymbol{\varepsilon}$ does not depend on $\lambda$ or $\sigma^{2}$)
\end{mycomment}

\begin{example2}
\label{BGI invar}Consider a balanced group interaction model with group fixed
effects. We have seen in Example \ref{exa GI failure} that in this model
Condition \ref{assum id} fails, because the columns of the fixed effects
matrix $I_{R}\otimes\iota_{m}$ span $\operatorname{col}(\omega_{\min}I_{n}%
-W)$. That is, in the notation introduced just before equation (\ref{Comegay}%
), $X_{\omega_{\min}}=I_{R}\otimes\iota_{m}$. Theorem \ref{theo non-identif}
therefore implies that, under Assumption \ref{assum distrib}, $(\lambda
,\beta,\sigma)$ cannot be identified from any statistic that is invariant
under $\mathcal{G}_{I_{R}\otimes\iota_{m}}^{1}$, even though the model is
invariant under that group (as shown in Example \ref{network FE}). Also,
recall from Example \ref{network FE} that reducing the data to $\mathcal{G}%
_{I_{R}\otimes\iota_{m}}^{1}$-invariant functions of the data removes the
group fixed effects. Thus, in this model, lack of identifiability from
$\mathcal{G}_{I_{R}\otimes\iota_{m}}^{1}$-invariant statistics corresponds to
the well-known identification failure that occurs upon removal of the group
fixed effects \citep{Lee2007b}.\hfill\qed

\end{example2}

Theorem \ref{theo non-identif} is connected to the results obtained in Section
\ref{sec ident}. Recall that the parameters of a network autoregression can be
identified by suitable restrictions on the variance structure of
$\boldsymbol{\varepsilon}$, regardless of whether Condition \ref{assum id}
holds; for example, this is certainly the case if $\mathrm{var}%
(\boldsymbol{\varepsilon})=I_{n}$, by Proposition \ref{lemma identif var}.
According to Theorem \ref{theo non-identif}, however, any result establishing
identification from the distribution of $\boldsymbol{y}$ is pointless when
Condition \ref{assum id} is not satisfied, because in that case the model is
invariant under the group $\mathcal{G}_{X_{\omega}}^{1}$, and yet
identification from $\mathcal{G}_{X_{\omega}}^{1}$-invariant functions of the
data is impossible. We illustrate this point in the context of Example
\ref{BGI invar}.

\begin{example2}
Consider the model in Example \ref{BGI invar}. Due to the failure of Condition
\ref{assum id}, any result establishing identification from the distribution
of $\boldsymbol{y}$ cannot help to achieve inference that respects the
invariance properties of the model. This is so, for example, for Proposition 2
in \cite{dePaula2017}, which establishes identification from the variance of
$\boldsymbol{y}$ for the particular case $R=1$, when $\left\vert
\lambda\right\vert <1$. While this identification result is correct, it should
be noted that inference based on it cannot respect the invariance properties
of the model. Indeed, the model is invariant under the group $\mathcal{G}%
_{\iota_{n}}^{1}$ of transformations $y\mapsto y+\alpha\iota_{n}$, $\alpha
\in\mathbb{R}$, and yet, by Theorem \ref{theo non-identif}, identification is
lost if, as advocated by the principle of invariance, we require inference to
satisfy the same symmetries. So, for instance, any invariant test in this
model can have only trivial power, and any equivariant estimator will be
useless.\hfill\qed

\begin{mycomment}
prin of invariance advocates invariant tests for invariant testing problems
\end{mycomment}

\begin{mycomment}
\begin{align*}
y  &  =\left(  I_{n}-\lambda B_{n}\right)  ^{-1}(\beta_{0}\iota_{n}+X^{\ast
}\beta^{\ast})+\sigma\left(  I_{n}-\lambda B_{m}\right)  ^{-1}\varepsilon\\
&  =\frac{1}{1-\lambda}\beta_{0}\iota_{n}+\left(  I_{n}-\lambda B_{m}\right)
^{-1}X^{\ast}\beta^{\ast}+\sigma\left(  I_{n}-\lambda B_{m}\right)
^{-1}\varepsilon
\end{align*}
\begin{align*}
g(y)  &  =\frac{1}{1-\lambda}\beta_{0}\iota_{n}+\alpha\iota_{n}+\left(
I_{n}-\lambda B_{m}\right)  ^{-1}X^{\ast}\beta^{\ast}+\sigma\left(
I_{n}-\lambda B_{m}\right)  ^{-1}\varepsilon\\
&  =\frac{1}{1-\lambda}\left(  \beta_{0}+(1-\lambda)\alpha\right)  \iota
_{n}+\left(  I_{n}-\lambda B_{m}\right)  ^{-1}X^{\ast}\beta^{\ast}%
+\sigma\left(  I_{n}-\lambda B_{m}\right)  ^{-1}\varepsilon
\end{align*}
induced group%
\[
\bar{g}\left(  \lambda,\beta_{0},\beta^{\ast},\sigma\right)  =\left(
\lambda,\beta_{0}+(1-\lambda)\alpha,\beta^{\ast},\sigma\right)
\]
say what equivariance would be in this context An estimator $\hat{\theta}$ of
a parameter $\theta$ based on data $y$ is said to be equivariant if
$\hat{\theta}(y+\alpha\iota_{n})=\bar{g}(\hat{\theta}(y))$
\end{mycomment}

\end{example2}

\begin{rem2}
\label{PrenPotscher remark}Theorem \ref{theo non-identif} is related to some
previous results in the literature. \cite{Martellosio2011} considers
non-identifiability from $G_{X}$-invariant statistics in a network
autoregression (\ref{SLM}) or spatial error model (\ref{SEM}) when $W$ is the
matrix $B_{n}\coloneqq\frac{1}{n-1}(\iota_{n}\iota_{n}^{\prime}-I_{n})$.
\cite{Preinerstorfer2015}, p. 30, generalizes \cite{Martellosio2011}'s results
to a regression model with correlated errors, which includes the particular
case of a spatial error model (\ref{SEM}) with arbitrary $W$. Neither of these
two papers, however, (i) discusses the relationship between
non-identifiability from invariant statistics and identifiability from the
first moment or from the first two moments; (ii) considers the set of
$\mathcal{G}_{X_{\omega}}^{1}$-invariant statistics, which is larger than the
set of $\mathcal{G}^{1}_{X}$-invariant statistics.
\end{rem2}

\begin{mycomment}
max inv under $\mathcal{G}_{X_{\omega}}^{1}$
$\mathcal{G}_{X}\coloneqq\{g_{\kappa,\delta}:\kappa>0,\delta\in\mathbb{R}%
^{k}\}$, where $g_{\kappa,\delta}$ denotes the function $y\mapsto\kappa
y+X\delta$ (a one-to-one transformation of $\mathbb{R}^{n}$), and its subgroup
$\mathcal{G}_{X}^{1}\coloneqq\{g_{1,\delta}:\delta\in\mathbb{R}^{k}\}$.
$r(y)=C_{X_{\omega}}y$
This can be established by direct verification of the definition of a maximal
invariant. For $\mathcal{G}_{X_{\omega}},$ $r(y)$ is invariant because
$r(y+X_{\omega}\delta)=v(y)$ for any $\delta\in\mathbb{R}^{...}$, since
$C_{X_{\omega}}X_{\omega}=0$, and it is different on different orbits because,
for any $y,\tilde{y}\in\mathbb{R}^{n}$, $r(y)=r(\tilde{y})$ if and only if
$C_{X_{\omega}}(\tilde{y}-y)=0$, which is equivalent to $\tilde{y}%
=y+X_{\omega}b$ for some $b\in\mathbb{R}^{...}$, thus proving maximality..
statistic invariant under $\mathcal{G}_{X}^{1}$ (i.e. depends on $y$ only
through $C_{X}y$) then it is invariant under $\mathcal{G}_{X_{\omega}}^{1}$
(i.e. depends on $y$ only through $C_{X_{\omega}}y$)
\end{mycomment}

\begin{mycomment}
removed as too cumbersome (but still interesting)
remark: At the end of Section \ref{sec inv} it was argued that Lemma
\ref{lemma identif var} cannot be useful for inference when Condition
\ref{assum id} fails. It is also worth pointing out that, when Condition
\ref{assum id} is satisfied, Proposition \ref{lemma identif var} is particularly
useful if $(\lambda,\beta)$ cannot be identified from the first moment. This
is the case for a network autoregression with $\mathrm{rank}(X,WX)=k$ (or for a network
error model). Examples of network autoregressions such that Condition \ref{assum id} is
satisfied and $\mathrm{rank}(X,WX)=k$ are given in Section \ref{sec id first}%
.\bigskip
\end{mycomment}

\begin{mycomment}
	of course in \ref{BGI invar} the same holds on replacing $\mathcal{G}_{I_{R}\otimes\iota_{m}}^{1}$ with $\mathcal{G}_{X}^{1}\supseteq
	\mathcal{G}_{I_{R}\otimes\iota_{m}}^{1}$
\end{mycomment}

\begin{mycomment}
	Some technical remarks on the results of this section can be found in Section \ref{sec suppl remarks}.
\end{mycomment}

\begin{mycomment}
	\emph{QUESTION}: How much of Propositions \ref{prop prof lik Assumption A} and
	\ref{prop identif} can be extended to regression model with general
	$\Sigma(\lambda)$?
	\begin{proposition}
		For $y=X\beta+u,$ if $C\Sigma(\lambda)C^{\prime}$ is a scalar multiple of
		$I_{n-k}$
		\begin{enumerate}
			\item[(i)] $l(\lambda)$ depends on y only through a term that does not depend
			on $\lambda.$
			???unbounded from above in a neighb of $a$
			\item[(ii)] $l_{\mathrm{a}}(\lambda)$ does not depend on $\lambda$
		\end{enumerate}
	\end{proposition}
	\begin{pff}%
		\begin{equation}
		l(\lambda)\coloneqq -\frac{n}{2}\log\left(  y^{\prime}U(\lambda)y\right)  -\frac{1}
		{2}\log\left(  \det\left(  \Sigma(\lambda)\right)  \right)  ,
		\end{equation}
		[remember that King's lemma says $U(\lambda)=C^{\prime}(C\Sigma(\lambda
		)C^{\prime})^{-1}C]$
		\begin{equation}
		l(\lambda)=-\frac{n}{2}\log\left(  y^{\prime}C^{\prime}(C\Sigma(\lambda
		)C^{\prime})^{-1}Cy\right)  -\frac{1}{2}\log\left(  \det\left(  \Sigma
		(\lambda)\right)  \right)  ,
		\end{equation}
		if $C\Sigma(\lambda)C^{\prime}=f(\lambda)I_{n-k}$,
		\begin{align*}
		l(\lambda)  &  \coloneqq -\frac{n}{2}\log\left(  f(\lambda)y^{\prime}M_{X}y\right)
		+\frac{1}{2}\log\left(  \det\left(  \Sigma^{-1}(\lambda)\right)  \right) \\
		&  =-\frac{n}{2}\log\left(  y^{\prime}M_{X}y\right)  -\frac{n}{2}\log\left(
		f(\lambda)\right)  +\frac{1}{2}\log\left(  \det\left(  \Sigma^{-1}
		(\lambda)\right)  \right)
		\end{align*}
		$\log\left(  \det\left(  \Sigma^{-1}(\lambda)\right)  \right)  \rightarrow
		-\infty$
		to establish lim$l(\lambda)$ we need to know $\lim f(\lambda)$
		Let $\Sigma^{-1}(\lambda)=DD^{\prime}$. under some conditions (or always?)
		$f(\lambda)=\lambda_{1}^{2}(D)$-------check this
		Letting $\lambda_{i}$ denote the distinct eigenvalues ($s$ and $n_{i}$ are
		indep of $\lambda$ except for isolated points)%
		\begin{align*}
		l(\lambda)  &  =-\frac{n}{2}\log\left(  f(\lambda)y^{\prime}M_{X}y\right)
		+\log\left(  \det\left(  DD^{\prime}\right)  \right)  =-\frac{n}{2}\log\left(
		y^{\prime}M_{X}y\right)  -\frac{n}{2}\log f(\lambda)+\log\left(  \det\left(
		D\right)  \right) \\
		&  =-\frac{n}{2}\log\left(  y^{\prime}M_{X}y\right)  -\frac{n}{2}\log\left(
		\lambda_{1}^{2}(D)\right)  +\log\left(  \prod_{i=1}^{s}\lambda_{i}^{n_{i}
		}(D)\right)
		\end{align*}
		careful as D not symmetric................
		I would not be surprised if what we need here is Ass 4 in PP!
		from PP: If B is a symmetric and nonnegative de\ldots nite l l matrix, every l
		l matrix A that satis\ldots es AA' = B is called a square root of B; with
		B$^{1/2}$ we denote its unique symmetric and nonnegative de\ldots nite square
		root. Note thatevery square root of B is of the form B$^{1/2}$U for some
		orthogonal matrix U.
		$\Sigma^{-1}(\lambda)=DD^{\prime}=\Sigma^{-1/2}(\lambda)UU^{\prime}
		\Sigma^{-1/2}(\lambda)$.............
		ii) yes this certainly holds for general $\Sigma(\lambda)$\bigskip
	\end{pff}
\end{mycomment}

\begin{mycomment}
	So, neither $l(\lambda)$ nor $l_{\mathrm{a}%
	}(\lambda)$ can provide a suitable basis for inference when Condition
	\ref{assum id} is violated.
\end{mycomment}

\begin{mycomment}
	Further properties of $l(\lambda)$ when Condition
	\ref{assum id} is violated are given by Lemma \ref{lemma l(lambda) violation}
	in the Supplement.
\end{mycomment}

\begin{mycomment}
	see potscher prein particularly remark 2.3%
\end{mycomment}

\begin{mycomment}
	OLD for lambda only:The profile log-likelihood functions of $\lambda$ in a SAR
	and in a network error model, given in equation (\ref{prof lik}) and equation
	(\ref{prof lik SEM}) in the Supplement, respectively, are identical if and
	only if $M_{S(\lambda)X}=M_{X}$, for any $\lambda$ such that $S(\lambda)$ is
	invertible. This occurs if and only if $\operatorname{col}(S(\lambda
	)X)=\operatorname{col}(X)$, for any $\lambda$ such that $S(\lambda)$ is
	invertible, which, as noted above, is equivalent to the condition that
	$\operatorname{col}(X)$ is an invariant subspace of $W$.
\end{mycomment}

\begin{mycomment}
	Note that $(\lambda,\sigma)$ is identifiable from $\mathrm{var}(y|W,X)$, via
	Proposition \ref{lemma identif var}. but inference based on $y$ is imposs - see
	\ref{prop prof lik Assumption A}
	Provided that $\mathrm{var}(\boldsymbol{\varepsilon}|W,X)$ does not depend on $\lambda$ or
	$\sigma$, the parameter $(\lambda,\sigma)$ is identified on $\Lambda
	\times(0,\infty)$, by Proposition \ref{lemma identif var}. However, since Condition
	\ref{assum id} fails for this model, $(\lambda,\sigma)$ cannot be identified
	from $\mathrm{var}(Cy|W,X)$.
\end{mycomment}

\subsection{\label{sec lik}Likelihood}

We now study the consequences of Theorem \ref{theo non-identif} for likelihood
inference. We consider the QMLE based on the original data, as introduced in
Section \ref{sec asy id}, but also the QMLE after transformation by
$C_{X_{\omega}}$ and the QMLE after transformation by $C_{X}$. Transformation
by $C_{X_{\omega}}$ is relevant, for example, when $X_{\omega}$ is a matrix of
fixed effects and one wishes to remove the fixed effects prior to estimation
\citep{Lee2007b,LeeLiuLin2010,LeeYu2010}. Transformation by $C_{X}$, on the
other hand, is relevant when the model is $\mathcal{G}_{X}^{1}$-invariant,
which is certainly the case when Condition \ref{assum id} fails. Note that,
when the model is $\mathcal{G}_{X}^{1}$-invariant, the QMLE after
transformation by $C_{X}$ is equivalent to the so-called adjusted QMLE, which
is obtained from the QMLE by centering the profile score for $(\lambda
,\sigma^{2})$ \citep[see][and Appendix \ref{app lik}]{Yu2015}. For estimation
of $(\lambda,\sigma^{2})$, it is well known that the adjusted QMLE usually
performs better than the QMLE when the dimension of $\beta$ is large with
respect to the sample size $n$ (including in fixed effects models, in which
case the dimension of $\beta$ is increasing with $n$).

We denote by $l(\lambda,\beta,\sigma^{2};y)$ the Gaussian quasi log-likelihood
for $(\lambda,\beta,\sigma^{2})$ in a network autoregression or in a network
error model, by $l(\lambda,\beta,\sigma^{2};Ay)$ the corresponding
log-likelihood obtained after premultiplying the model by a matrix $A$, and by
$l(\lambda;y)$ the profile likelihood for $\lambda$. The next result shows
that, when Condition \ref{assum id} fails, likelihood estimation is fruitless,
before or after transformation of the data.

\begin{mycomment}
could alternatively express the prop in terms of $l(\lambda,\sigma^{2})$ and $l_{\mathrm{a}}		(\lambda,\sigma^{2})$ (the latter corresponds to density max inv under $G_^1_X$)
\end{mycomment}

\begin{mycomment}
	unnique max of limiting quasi lik is nec and suff for identification from the quasi lik - see Lee and Yu 2015 and cf Remark \ref{rem id}
\end{mycomment}

\begin{mycomment}
	The reverse implication $\operatorname{col}(X)$ is an
	invariant subspace of $W$ then Condition \ref{assum id} is violated does not hold. For
	example, consider the group interaction weights matrix of Example \ref{exa GI} (UGI)
	and the fixed effects matrix $X=\bigoplus_{i=1}^{R}\iota_{m_{i}}$. In that
	case, $\operatorname{col}(X)$ is an invariant subspace of $W$ for any $R$ and
	for any $m_{1},\ldots,m_{R}$, but Condition \ref{assum id} is violated (with
	$\omega=\omega_{\min}$) only if the model is balanced (i.e. $m_{1}=\ldots=m_{r}$).
\end{mycomment}

\begin{proposition}
\label{prop prof lik Assumption A}Consider the network autoregression
(\ref{SLM}) or the network error model (\ref{SEM}), with $\lambda$ such that
$\det(S(\lambda))\neq0$, and $y\notin\operatorname{col}(X)$. If Condition
\ref{assum id} is violated for some eigenvalue $\omega$ of $W$, then:

\begin{enumerate}
\item[(i)] the log-likelihood functions $l(\lambda,\beta,\sigma^{2}%
;C_{X_{\omega}}y)$ and $l(\lambda,\beta,\sigma^{2};C_{X}y)$ do not depend on
$(\lambda,\beta,\sigma^{2})$;

\item[(ii)] the profile score associated with the profile log-likelihood
function $l(\lambda;y)$ does not depend on $y\ $and $X$.
\end{enumerate}
\end{proposition}

Part (i) of Proposition \ref{prop prof lik Assumption A} says that the
functions $l(\lambda,\beta,\sigma^{2};C_{X_{\omega}}y)$ and $l(\lambda
,\beta,\sigma^{2};C_{X}y)$ are constant. by Proposition
\ref{lemma identif var} this cannot be the case for the likelihood based on
$y$, $l(\lambda,\beta,\sigma^{2};y)$. However, part (ii) of Proposition
\ref{prop prof lik Assumption A} establishes that the first derivative of the
profile likelihood $l(\lambda;y)$ does not depend on the data, and
consequently that a QMLE based on $l(\lambda;y)$ cannot depend on the data if
it exists.\footnote{\label{foot KPZ}This is similar to a result contained in
Theorem 1 of \cite{KelejianPruchaYuzefovich2006}. That result establishes
that, in a network autoregression with $W=B_{n}$, the 2SLS estimator of
$\lambda$ considered there does not depend on the data.} This results can be
understood in terms of the invariance results in Section \ref{sec role}: if
Condition \ref{assum id} is violated for an eigenvalue $\omega$ of $W$, then
$l(\lambda;y)$ is a $\mathcal{G}_{X_{\omega}}^{1}$-invariant loss function
(see equation (\ref{prof lik ASSUM1 violated}) in the proof of the
proposition), and as such it cannot produce a useful estimator.

\begin{mycomment}
REVISEEEE It is easily verified that the maximal invariant under the group
induced by $\mathcal{G}_{X}$ on the parameter space is $\lambda$ (it would be
$\left(  \lambda,\eta\right)  $ in the presence of a parameter $\eta$ in the
distribution of $\boldsymbol{\varepsilon}$). This may seem to contradict one
of the fundamental results on invariance, which is usually stated by saying
that the distribution of an invariant statistic depends \textit{only} on a
maximal invariant induced on the parameter space \citep[e.g.,][Theorem
	6.3.2]{Lehmann2005}. The apparent contradiction is due to the
non-identification caused by the violation of Condition \ref{assum id}%
\end{mycomment}

\begin{mycomment}
This result is a consequence of the fact that the only part of the profile
log-likelihood function $l(\lambda;y)$ that depends on the data $y\ $and $X$
does not depend on $\lambda$;$\ $see equation (\ref{prof lik ASSUM1 violated})
in the proof of the proposition.
\end{mycomment}

\begin{example2}
\label{exa network FE ML}We have seen in Example \ref{network FE} that, for a
network fixed effects model in which all interaction matrices $W_{r}$ are
row-stochastic, transformation by $C_{X_{\mathrm{FE}}}$ is equivalent to the
transformation proposed in \cite{LeeLiuLin2010}. Thus the QMLE based on the
Gaussian likelihood $l(\lambda,\beta,\sigma^{2};C_{X_{\omega}}y)$ is
equivalent to the QMLE considered in \cite{LeeLiuLin2010}. In the particular
case of a balanced group interaction model (i.e., $W_{r}=B_{m_{r}}$, for each
$r=1,\ldots,R$, and $m_{1}=m_{2}=\ldots= m_{R}$; see Example \ref{exa GI},
Condition \ref{assum id} fails, with $X_{\omega}=X_{\mathrm{FE}}$; see Example
\ref{exa GI failure}. Thus, for the balanced group interaction model,
Proposition \ref{prop prof lik Assumption A} establishes that the Gaussian
quasi-likelihood based $C_{X_{\mathrm{FE}}}y$ and that based on $C_{X}y$ are
constant functions of the parameters.
\end{example2}

\begin{mycomment}
	note MLE is equiv under group under which model is inv. adj MLE is inv
\end{mycomment}

\begin{mycomment}
	............Consider the network autoregression (\ref{SLM}) with $W$ and $X$ such that
	Condition \ref{assum id} fails, and suppose Assumption \ref{assum distrib}
	holds. Recall that the model is $\mathcal{G}_{X}$-invariant by Lemma
	\ref{lemma model inv}, and that a maximal invariant under $\mathcal{G}_{X}$ is
	$v\coloneqq C_{X}y/\left\Vert C_{X}y\right\Vert $. Since $\sigma
	/(1-\lambda\omega)$ appears as a scale factor in \eqref{Cy eq}, it follows
	that the distribution of $v$, and hence of any $\mathcal{G}_{X}$-invariant
	statistic, is free of $\lambda$ if the distribution of $\boldsymbol{\varepsilon}$ does not depend on $\lambda$ (and of
	course is also free of $\sigma^{2}$ and $\beta$ under the full Assumption
	\ref{assum distrib}). It is easily verified that the maximal invariant induced on the parameter
	space is $\lambda$. This might seem to contradict one of the fundamental
	results on invariance, which is usually stated by saying that the distribution
	of an invariant statistic depends \textit{only} on a maximal invariant induced
	on the parameter space \citep[e.g.,][Theorem
	6.3.2]{Lehmann2005}. The apparent contradiction is due to the
	non-identifiability caused by the violation of Condition \ref{assum id}%
	.\textcolor{red}{same applies to SEM}
\end{mycomment}

\begin{mycomment}
	We say that $\lambda$ is identifiable from the quasi profile likelihood
	$l(\lambda)$ if $l(\lambda_{1})=l(\lambda_{2})$ for almost all $y\in
	\mathbb{R}^{n}$. Of course, identifiability of $\lambda$ from $\mathrm{E}%
	(y|W,X)$ (Proposition \ref{lemma identif mean}) or from $\mathrm{var}(y|W,X)$ (Lemma
	\ref{lemma identif var}) implies identifiability from $l(\lambda)$.
\end{mycomment}

\begin{mycomment}
	for the fact that the eigenvalues of $A$ are eigenvalues of $W$ see
	$\backslash$%
	citep[e.g.,][Theorem 3.9]\{StewartSun1990\}[[[[[[[[[[no need for reference
	this is straightforward]]]]]]]]]]
\end{mycomment}

\begin{mycomment}
	This not need anymore as I moved assumption: Proposition \ref{lemma identif mean} does not require the assumption, made in
	Section \ref{sec SLM}, that $W$ has at least one negative eigenvalue and at
	least one positive eigenvalue.
\end{mycomment}

\section{\label{Sec simul}Simulations}

Section \ref{sec id first} contains several examples in which $\mathrm{rank}%
(X,WX)=k$, and therefore $\lambda$ and $\beta$ cannot be identified from the
first moment of $\boldsymbol{y}$, that is, cannot be identified when the only
assumption about $\boldsymbol{\varepsilon}$ is $\mathrm{E}%
(\boldsymbol{\varepsilon})=0$. As noted in that section, however, the
condition $\mathrm{rank}(X,WX)=k$ is very strong in general. What might be
more relevant in typical applications is that the condition is close, in some
sense, to being satisfied. In that case, it would be natural to expect that
identification from the first moment will be weak. This section analyses, by
simulation, the consequences of near non-identification. We consider two Monte
Carlo experiments, designed to study what happens close to, respectively, (i)
a case when Condition 1 holds but $\mathrm{rank}(X,WX)=k$, (ii) a case when
Condition 1 fails.

In both experiments, we draw $10,000$ replications from model (\ref{SLM}),
with errors drawn from either a standard normal distribution or a gamma
distribution with shape parameter 1 and scale parameter 1, demeaned by the
population mean. Mean, variance, skewness, and kurtosis are 0, 1, 0, and 3 for
the former distribution and 0, 1, 2, and 9 for the latter. The main objective
of the simulations is to study the behavior of an estimator that uses only
first moment information. We focus on the two-stage least squares estimator
(2SLSE) with instruments for $(X,Wy)$ given by (the linearly independent
columns of) $(X,WX,W^{2}X)$ \citep{Kelejian98}. We study two implications of
near non-identification: (i) accuracy of the estimator; (ii) adequacy of
first-order asymptotic approximation to the distribution of the estimator.
Accuracy of the estimator is measured by the median square error. The root
median square error is reported rather than the more usual root mean square
error because, in the setting we are considering, the variance of the 2SLS
estimator does not exist \citep[see][Section 7.2.2]{Roberts95}. Adequacy of
the first-order asymptotic approximation is measured by the coverage of 95\%
Wald confidence intervals.\footnote{For a parameter $\phi$, the 95\%\ Wald
confidence interval is $\hat{\phi}\pm1.96\sqrt{\hat{v}}$, where $\hat{\phi}$
is an estimator of $\phi$ and $\hat{v}$ its estimated asymptotic variance.
Expressions for the asymptotic variance of the 2SLS is standard, and that of
the QMLE is given in \cite{Lee2004}, Theorem 3.2.} As a benchmark, we consider
the (quasi) maximum likelihood estimator based on the Gaussian likelihood,
abbreviated by (Q)MLE; see Section \ref{sec lik}). Contrary to the 2SLSE, the
QMLE also uses second moment information.

\subsection{\label{sec dgp1}First experiment}

In the first experiment, $n$ is either $100$ or $1000$, and $W$ is a
row-normalized 2-ahead 2-behind interaction matrix (before row-normalization,
this is a matrix with all entries in the two diagonals above and the two
diagonals below the main diagonal equal to one, and zero everywhere else), and
the model has a single regressor equal to $\iota_{n}+bz$, where $b\in
\mathbb{R}$ and $z\thicksim\mathrm{N}(0,I_{n})$, with $z$ being generated
once, for each $n$, and then kept fixed across replications. If $b=0$, then
$\mathrm{rank}(X,WX)=k=1$ (see Example \ref{exa netwFE}(ii), with $R=1$), and
therefore the parameters $\lambda$ and $\beta$ cannot be identified from the
first moment. Thus, we expect any estimator of $\lambda$ and $\beta$ that
relies entirely on the specification of the first moment of $\boldsymbol{y}$
to perform poorly if $b$ is close to $0$ (and to be undefined when $b=0$). The
true values of $\lambda$, $\beta$, $\sigma$ are set to $\lambda_{0}=0$,
$\beta_{0}=0.1,1$, and $\sigma_{0}=1$. Table \ref{tab:RMdSE} displays the root
median square error of the 2SLSE and (Q)MLE of $\lambda$ and $\beta$. Dots
indicate nonexistence of the estimator. The 2SLSE exploits the (correct)
specification of the first moment, whereas the QMLE also exploits the
(correct, given that the data are generated under the assumption
$\operatorname{var}(\boldsymbol{\varepsilon})=I_{n}$) specification of the
second moment. Let us look at the case $\beta_{0}=1$ first. For both $\lambda$
and $\beta$, and for both the normal and the gamma distributions, the
performance of the 2SLSE is satisfactory, compared to the (Q)MLE benchmark,
when $b=1$, but deteriorates rapidly as $b$ gets smaller. Such a deterioration
is due to both the bias and the dispersion of the 2SLSE growing large as $b$
decreases, for any $n$. When $b=0,$ the 2SLSE is not defined. On the contrary,
due to the fact that it also exploits second moment information, the (Q)MLE is
not much affected by the lack of identifiability from the first moment that
occurs when $b=0$. Indeed, the root median square error of the (Q)MLE is
considerably less sensitive to $b$, and the (Q)MLE does well even when $b=0$.
Moving to the case $\beta_{0}=0.1$, it is natural to expect that
identifiability from the fist moment may become more difficult when $\beta$ is
near zero. Indeed, all values of $(\lambda,\beta)$ in the $\mu_{\mathbb{R}%
^{2}}$-null set $\Lambda_{\mathrm{u}}\times\{0\}$ cannot be identified from
the first moment --- see case (a) in the proof of Proposition 3.1. In
particular, the Monte Carlo results show that the 2SLE of $\lambda$ performs
much worse than the (Q)MLE, even when $b=1$.

Table \ref{tab:RMdSE} confirms that estimation based on the first moment is
impossible when $\mathrm{rank}(X,WX)=k$ and difficult in cases close to
$\mathrm{rank}(X,WX)=k$. A second consequence of $X$ and $W$ being such that
$\mathrm{rank}(X,WX)$ is close to being equal to $k$ is that typical
asymptotic approximations may become unreliable. Table \ref{tab:cover}
displays coverages of 95\% two-sided Wald confidence intervals based on
asymptotic normality, in the same setting at Table \ref{tab:RMdSE}. When
$\beta_{0}=1$, the empirical coverages for the 2SLSE are close to the nominal
one when $b=1$, but get further and further away as $b$ decreases. When
$\beta_{0}=0.1,$ the empirical coverages for the 2SLSE are poor even when
$b=1$. The (Q)MLE, on the other hand does well in terms of coverage even when
$b=0$ and even when $\beta_{0}=0.1$, again due to the fact that, in these
simulations, it does not rely on the specification of the first moment only,
but also exploits the (correct in these simulations) specification of the
second moment.\vspace{.7cm} \begin{mycomment}
	I haven;t used the sandwic yet for gamma but this should not make much difference
\end{mycomment}
\begin{table}[th!]
\caption{Root median square error of the 2SLS and (Q)ML estimators of
$\lambda$ and $\beta$ in the first experiment.}%
\label{tab:RMdSE}
\centering
\begin{adjustbox}{width=1\textwidth,center=\textwidth}
		
		\begin{tabular}[c]{m{.8cm}m{1cm}m{.9cm}m{1mm}m{1.3cm}m{.8cm}m{5mm}m{1.3cm}m{.8cm}m{5mm}m{1.3cm}m{.8cm}m{5mm}m{1.3cm}m{.8cm}}
			\hline
			&&   & & \multicolumn{2}{l}{\text{Normal}}    &  && && \multicolumn{2}{l}{\text{Gamma}}\bigstrut[t]     &  &&   \\
			\cline{5-9}\cline{11-15}
			&&      & & $\lambda$                   &                           &  & $\beta$       &            &                      & $\lambda$                   &                           &  & $\beta$\bigstrut[t] &                         \\
			\cline{5-6}\cline{8-9}\cline{11-12}\cline{14-15}	
			$\beta_0$ &$n$       & $b$    &  &\text{2SLS}&\text{ML}&&\text{2SLS}&\text{ML}&&\text{2SLS}&\text{QML}&&\text{2SLS}&\text{QML}\bigstrut[t]  \\ \hline
\\[-6mm]\\
			1&100 	& 1    &  & 0.069 & 0.056 &  & 0.055       & 0.051      &  & 0.068 & 0.055 &  & 0.054 & 0.050                  \\
			&& 0.1  &  & 0.489 & 0.094 &  & 0.481       & 0.115      &  & 0.479 & 0.092 &  & 0.476 & 0.110                   \\
			&& 0.01 &  & 1.464 & 0.096 &  & 1.451       & 0.119      &  & 1.472 & 0.093 &  & 1.451 & 0.114                   \\
			&& 0 &  & \multicolumn{1}{c}{$\sbullet$} & 0.096 &  &\multicolumn{1}{c}{$\sbullet$}       &  0.120     &  &\multicolumn{1}{c}{$\sbullet$}  & 0.093&  & \multicolumn{1}{c}{$\sbullet$} &  0.114 \bigstrut[b]                 \\
			&1000 	& 1    &  & 0.025 & 0.019 &  & 0.020       & 0.018      &  & 0.026 & 0.020 &  & 0.020 & 0.018\bigstrut[t]                   \\
			&& 0.1  &  & 0.195 & 0.030 &  & 0.194       & 0.037      &  & 0.195 & 0.030 &  & 0.194 & 0.037                   \\
			&& 0.01 &  & 1.183 & 0.031 &  & 1.183       & 0.037      &  & 1.199 & 0.030 &  & 1.194 & 0.037                   \\
			&& 0 &  & \multicolumn{1}{c}{$\sbullet$} & 0.031 &  &\multicolumn{1}{c}{$\sbullet$}       &  0.037     &  &\multicolumn{1}{c}{$\sbullet$}  & 0.030&  & \multicolumn{1}{c}{$\sbullet$} &  0.037                  \\
\\[-6mm]\\
			0.1&100 & 1    &  & 0.617	 & 0.092 &  & 0.056 & 0.042     &  & 0.616	& 0.092 &  & 0.055	 & 0.042                  \\
			&		& 0.1  &  & 1.445	 & 0.095 &  & 0.149 & 0.068     &  & 1.449	& 0.093 &  &  0.147	 & 0.066                   \\
			&		& 0.01 &  & 1.596	 & 0.096 &  & 0.159	& 0.070     &  & 1.593	& 0.093 &  & 0.160	 & 0.067                  \\
			&		& 0 &  & \multicolumn{1}{c}{$\sbullet$} & 0.096 &  &\multicolumn{1}{c}{$\sbullet$}   &  0.070     &  &\multicolumn{1}{c}{$\sbullet$}  & 0.093&  & \multicolumn{1}{c}{$\sbullet$} &  0.067 \bigstrut[b]                 \\
			&1000 	& 1    &  & 0.248	 & 0.030 &  & 0.020	 & 0.016    &  & 0.252	& 0.030 &  & 0.020	& 0.015 \bigstrut[t]                   \\
			&		& 0.1  &  &  1.190	 & 0.030 &  & 0.111	 & 0.022    &  & 1.205	& 0.030	 &  & 0.112	 & 0.021                   \\
			&		& 0.01 &  &  1.623	 & 0.030 &  & 0.144	 & 0.022    &  & 1.669	& 0.030 &  & 0.150	 & 0.021                   \\
			&		& 0 &  & \multicolumn{1}{c}{$\sbullet$} & 0.031 &  &\multicolumn{1}{c}{$\sbullet$}       &  0.022     &  &\multicolumn{1}{c}{$\sbullet$}  & 0.030&  & \multicolumn{1}{c}{$\sbullet$} &  0.021                  \\
			\hline
		\end{tabular}
	\end{adjustbox}
\vspace{.6cm}
\caption{Coverage of 95\% confidence intervals for $\lambda$ and $\beta$ in
	the first experiment.\vspace{.0cm}}%
\label{tab:cover}
\begin{adjustbox}{width=1\textwidth,center=\textwidth}
	
	\begin{tabular}[c]{m{.8cm}m{1cm}m{.9cm}m{1mm}m{1.3cm}m{.8cm}m{5mm}m{1.3cm}m{.8cm}m{5mm}m{1.3cm}m{.8cm}m{5mm}m{1.3cm}m{.8cm}}
		\hline
		&&   & & \multicolumn{2}{l}{\text{Normal}}    &  && && \multicolumn{2}{l}{\text{Gamma}}\bigstrut[t]     &  &&   \\
		\cline{5-9}\cline{11-15}
		&&      & & $\lambda$                   &                           &  & $\beta$       &            &                      & $\lambda$                   &                           &  & $\beta$\bigstrut[t] &                         \\
		\cline{5-6}\cline{8-9}\cline{11-12}\cline{14-15}	
		$\beta_0$&$n$       & $b$    &  &\text{2SLS}&\text{ML}&&\text{2SLS}&\text{ML}&&\text{2SLS}&\text{QML}&&\text{2SLS}&\text{QML}\bigstrut[t]  \\ \hline
		\\[-6mm]\\
		1&100 & 1 && 0.947 & 0.945 &  & 0.948 & 0.947 && 0.949 & 0.944 && 0.946 & 0.944\\
		&& 0.1 && 0.985 & 0.945 &  & 0.985 & 0.944 && 0.983 & 0.950 && 0.982 & 0.947\\
		&& 0.01 && 0.995 & 0.943 &  & 0.995 & 0.942&& 0.994& 0.951&& 0.994&0.950\\
		&& 0 &  & \multicolumn{1}{c}{$\sbullet$} & 0.943 &&\multicolumn{1}{c}{$\sbullet$}       & 0.942 &&\multicolumn{1}{c}{$\sbullet$}  & 0.952&& \multicolumn{1}{c}{$\sbullet$} & 0.951   \bigstrut[b]                 \\
		&1000 	& 1    && 0.950 & 0.948 &  & 0.950 & 0.950 && 0.951&0.951 && 0.949&0.953 \bigstrut[t]\\
		&& 0.1 && 0.962 & 0.953 &  & 0.963 & 0.952 && 0.961&0.951 && 0.961&0.951\\
		&& 0.01 && 0.995 & 0.951 &  & 0.995 & 0.952&& 0.994&0.951 && 0.995&0.950\\
		&& 0 &  & \multicolumn{1}{c}{$\sbullet$} & 0.951 &&\multicolumn{1}{c}{$\sbullet$}       & 0.952 &&\multicolumn{1}{c}{$\sbullet$}  & 0.951 && \multicolumn{1}{c}{$\sbullet$} &  0.950 \\
		\\[-6mm]\\
		0.1&100 & 1    &  & 0.986	 & 0.946 &  & 0.977	& 0.945      &  & 0.990	 & 0.951 &  &  0.977	& 0.937                  \\
		&		& 0.1  &  & 0.995	 & 0.943 &  & 0.994	& 0.943	      &  & 0.994	 & 0.951 &  & 0.990	& 0.936                  \\
		&		& 0.01 &  & 0.995	 & 0.943 &  & 0.994	& 0.944      &  & 0.994	 & 0.952 &  &  0.992	& 0.938                  \\
		&		& 0 &  & \multicolumn{1}{c}{$\sbullet$} & 0.943 &  &\multicolumn{1}{c}{$\sbullet$}       &  0.944     &  &\multicolumn{1}{c}{$\sbullet$}  & 0.952&  & \multicolumn{1}{c}{$\sbullet$} & 0.938 \bigstrut[b]                 \\
		&1000 	& 1    &  & 0.971	 & 0.953 &  & 0.962	& 0.949      &  & 0.969	 & 0.951 &  & 0.960	& 0.950	 \bigstrut[t]                   \\
		&		& 0.1  &  & 0.995	 & 0.951 &  &  0.994	& 0.950      &  & 0.995	 & 0.951 &  & 0.994	& 0.947                 \\
		&		& 0.01 &  & 0.995	 & 0.951 &  &  0.995	& 0.950      &  & 0.996	 & 0.951 &  & 0.995	& 0.949                   \\
		&		& 0 &  & \multicolumn{1}{c}{$\sbullet$} & 0.951 &  &\multicolumn{1}{c}{$\sbullet$}       &  0.951     &  &\multicolumn{1}{c}{$\sbullet$}  & 0.951&  & \multicolumn{1}{c}{$\sbullet$} &  0.949                  \\
		\hline
	\end{tabular}
	
\end{adjustbox}

\end{table}
\subsection{\label{sec dgp2}Second experiment}

The second data generating process is the Group Interaction model of Example
\ref{exa GI} with fixed effects. The model equation is%
\begin{equation}
\boldsymbol{y}_{r}=\lambda B_{m_{r}}\boldsymbol{y}_{r}+\tilde{\beta}
\widetilde{x}_{r}+\boldsymbol{\alpha}_{r}\iota_{m_{r}}+\sigma
\boldsymbol{\varepsilon}_{r},\text{ }r=1,\ldots,R. \label{exp 2}%
\end{equation}

The number of groups is $R=50,100,200,$ and for the group sizes we consider 6
cases, all with average group size equal to 10 (so that, corresponding to
$R=50,100,200$, we have $n=500,1000,2000$), but with various degrees of
unbalancedness. Specifically, the sequence of group sizes $m_{1},\ldots,m_{R}$
is periodic, with period 10 (i.e., $m_{i}=m_{i+10}$, for any $i=1,\ldots
,R-10$), and the first $10$ group sizes $m_{1},\ldots,m_{10}$ are given in
Table \ref{tab:group sizes}. For each $R$, the single regressor $\widetilde{x}%
=(\widetilde{x}_{1}^{\prime},\ldots,\widetilde{x}_{R}^{\prime})^{\prime}$ is
drawn once from $\mathrm{N}(0,I_{n})$ and then kept fixed across replications,
and $\boldsymbol{\alpha}=(\boldsymbol{\alpha}_{1},\ldots,\boldsymbol{\alpha
}_{R})^{\prime}$ is drawn from $\mathrm{N}(0,I_{R})$ in each replication. The
true values of $\lambda$, $\beta$, $\sigma$ are set to $\lambda_{0}=0$,
$\tilde{\beta}_{0}=1$, and $\sigma_{0}=1$.

\begin{table}[h]
\caption{Group sizes in the second experiment.}%
\label{tab:group sizes}%
\centering
\begin{adjustbox}{width=.65\textwidth,center=\textwidth}
		$	
		\begin{tabular}[l]{cccccccccccc}%
			\hline
			\text{case} &  & $m_{1}$ & $m_{2}$ & $m_{3}$ & $m_{4}$ & $m_{5}$ & $m_{6}$ & $m_{7}$ & $m_{8}$ & $m_{9}$ & $m_{10}$\bigstrut[t]\bigstrut[b]\\
			\hline
			1 &  & 2 & 4 & 6 & 8 & 9 & 10 & 12 & 14 & 16 & 18\bigstrut[t]\\
			2 &  & 4 & 6 & 8 & 10 & 10 & 10 & 10 & 12 & 14 & 16\\
			3 &  & 6 & 8 & 10 & 10 & 10 & 10 & 10 & 10 & 12 & 14\\
			4 &  & 8 & 10 & 10 & 10 & 10 & 10 & 10 & 10 & 10 & 12\\
			5 &  & 9 & 10 & 10 & 10 & 10 & 10 & 10 & 10 & 10 & 11\\
			6 &  & 10 & 10 & 10 & 10 & 10 & 10 & 10 & 10 & 10 & 10\\\hline
		\end{tabular}
		$
	\end{adjustbox}
\end{table}

Estimation is performed after removal of the fixed effects by
premultiplication by $C_{X_{\mathrm{FE}}}$ (see Section \ref{sec lik}%
).\footnote{Note that the 2SLSE of $(\lambda,\tilde{\beta} )$ after
premultiplication by $C_{X_{\mathrm{FE}}}$ is the same as the 2SLSE prior to
removal of the fixed effects.} When the model is balanced (case 6), both the
2SLSE and the (Q)MLE do not exist. The 2SLSE does not exist because
$\mathrm{rank}(X,WX)=k=1$ and therefore the instrument matrix is singular. The
(Q)MLE does not exist because Condition \ref{assum id} fails and therefore the
likelihood based on $C_{X_{\mathrm{FE}}}\boldsymbol{y}$ is constant (see
Example \ref{exa network FE ML}). Table \ref{tab:RMdSE2} shows that the root
median square error of both 2SLS and (Q)ML estimators increases as the model
becomes more balanced. Similarly to the first experiment, it is not surprising
that the (Q)MLE performs better than the 2SLSE, given that $\operatorname{var}%
(\boldsymbol{\varepsilon})=I_{n}$ in the DGP. It is worth noting, however,
that in the current experiment, the root median square error of the (Q)MLE,
not only that of the 2SLSE, increases substantially close to the
non-identifiability case. This is due to the two different types of
non-identifiability studied in the two experiments.

Table \ref{tab:cover2} shows that coverages of Wald confidence intervals can
be very far from the nominal coverage when the model is close to being
balanced. This is true for both the 2SLSE and (Q)MLE.

\begin{table}[t!]
\caption{Root median square error of the 2SLS and (Q)ML estimators of
$\lambda$ and $\tilde{\beta} $ in the second experiment.}%
\label{tab:RMdSE2}
\centering
\begin{adjustbox}{width=1\textwidth,center=\textwidth}
			
		\begin{tabular}[c]{m{1cm}cm{3mm}m{1.3cm}m{.8cm}m{5mm}m{1.3cm}m{.8cm}m{5mm}m{1.3cm}m{.8cm}m{5mm}m{1.3cm}m{.8cm}}
			\hline
			&   & & \multicolumn{2}{l}{\text{Normal}}    &  && && \multicolumn{2}{l}{\text{Gamma}}\bigstrut[t]     &  &&   \\
			\cline{4-8}\cline{10-14}\\[-4mm]
			&      & & $\lambda$                   &                           &  & $\tilde{\beta}
			$       &            &                      & $\lambda$                   &                           &  & $\tilde{\beta}
			$\bigstrut[t] &                         \\ \cline{4-5}\cline{7-8}\cline{10-11}\cline{13-14}	
			$R$       & case    &  &\text{2SLS}&\text{ML}&&\text{2SLS}&\text{ML}&&\text{2SLS}&\text{QML}&&\text{2SLS}&\text{QML}\bigstrut[t]  \\ \hline
			50 	& 1 && 0.227 & 0.152 && 0.044	& 0.037 && 0.215 & 0.216 && 0.043 & 0.041\\
			& 2 && 0.499 & 0.328 && 0.068	& 0.052 && 0.455& 0.465 && 0.068 & 0.063\\
			& 3 && 0.999 & 0.662 && 0.121	& 0.085 && 0.917 & 0.921 && 0.110 & 0.113\\
			& 4 && 2.403 & 1.453 && 0.273	& 0.170 && 2.323 & 2.146 && 0.267 & 0.248\\
			& 5 && 4.820 & 2.917 && 0.538	& 0.328 && 4.786 & 4.092 && 0.535 & 0.462\\
			& 6 && \multicolumn{1}{c}{$\sbullet$} & \multicolumn{1}{c}{$\sbullet$} && \multicolumn{1}{c}{$\sbullet$} & \multicolumn{1}{c}{$\sbullet$} && \multicolumn{1}{c}{$\sbullet$} & \multicolumn{1}{c}{$\sbullet$} && \multicolumn{1}{c}{$\sbullet$} &  \multicolumn{1}{c}{$\sbullet$} \bigstrut[b]                 \\
			100 & 1 && 0.152 & 0.106 &&  0.029	& 0.026 && 0.149 & 0.141 && 0.030 & 0.029\bigstrut[t]\\
			& 2 && 0.387 & 0.243 &&  0.050	& 0.036 && 0.327 & 0.332 && 0.046 & 0.046\\
			& 3 && 0.735 & 0.460 &&  0.087	& 0.056 && 0.702 & 0.697 && 0.084 & 0.084\\
			& 4 && 1.755 & 1.110 &&  0.198	& 0.123 && 1.726 & 1.641 && 0.197 & 0.183\\
			& 5 && 3.568 & 2.134 &&  0.398	& 0.241 && 3.563 & 3.290 &&  0.397 &  0.370\\
			& 6 && \multicolumn{1}{c}{$\sbullet$} & \multicolumn{1}{c}{$\sbullet$} && \multicolumn{1}{c}{$\sbullet$} & \multicolumn{1}{c}{$\sbullet$} && \multicolumn{1}{c}{$\sbullet$} & \multicolumn{1}{c}{$\sbullet$} && \multicolumn{1}{c}{$\sbullet$} &  \multicolumn{1}{c}{$\sbullet$} \bigstrut[b]                 \\
			200 & 1 && 0.111 & 0.079 && 0.020	& 0.018  && 0.102 & 0.097 && 0.021 & 0.020\bigstrut[t]\\
			& 2 && 0.276 & 0.174 && 0.035	& 0.026  && 0.272 & 0.261 &&  0.036 &  0.035\\
			& 3 && 0.551 & 0.317 && 0.064	& 0.040  && 0.526 & 0.507 && 0.060 & 0.061\\
			& 4 && 1.301 & 0.795 && 0.145	& 0.088  && 1.292 & 1.220 && 0.144 & 0.137\\
			& 5 && 2.517 & 1.595 && 0.282	& 0.179  && 2.585 & 2.400 &&  0.290 &  0.290\\
			& 6 && \multicolumn{1}{c}{$\sbullet$} & \multicolumn{1}{c}{$\sbullet$} && \multicolumn{1}{c}{$\sbullet$} & \multicolumn{1}{c}{$\sbullet$} && \multicolumn{1}{c}{$\sbullet$} & \multicolumn{1}{c}{$\sbullet$} && \multicolumn{1}{c}{$\sbullet$} &  \multicolumn{1}{c}{$\sbullet$} \bigstrut[b]                 \\
			\hline
		\end{tabular}
		
	\end{adjustbox}
\vspace{.6cm}
\caption{Coverage of 95\% confidence intervals for $\lambda$ and $\tilde
	{\beta}$ in the second experiment.\vspace{.0cm}}%
\label{tab:cover2}%
\begin{adjustbox}{width=1\textwidth,center=\textwidth}
	
	\begin{tabular}[c]{m{1cm}cm{3mm}m{1.3cm}m{.8cm}m{5mm}m{1.3cm}m{.8cm}m{5mm}m{1.3cm}m{.8cm}m{5mm}m{1.3cm}m{.8cm}}
		\hline
		&   & & \multicolumn{2}{l}{\text{Normal}}    &  && && \multicolumn{2}{l}{\text{Gamma}}\bigstrut[t]     &  &&   \\
		\cline{4-8}\cline{10-14}\\[-4mm]
		&      & & $\lambda$                   &                           &  & $\tilde{\beta}
		$       &            &                      & $\lambda$                   &                           &  & $\tilde{\beta}
		$\bigstrut[t] &                         \\ \cline{4-5}\cline{7-8}\cline{10-11}\cline{13-14}	
		$R$       & case    &  &\text{2SLS}&\text{ML}&&\text{2SLS}&\text{ML}&&\text{2SLS}&\text{QML}&&\text{2SLS}&\text{QML}\bigstrut[t]  \\ \hline
		50 	& 1 && 0.960 & 0.959 && 0.950 & 0.954 && 0.941 & 0.873 && 0.951 & 0.925\\
		& 2 && 0.929 & 0.947 && 0.935	& 0.949 && 0.945 & 0.835 && 0.933 & 0.844\\
		& 3 && 0.912 & 0.951 && 0.917	& 0.953 && 0.920 & 0.856 && 0.923 & 0.869\\
		& 4 && 0.790 & 0.928 && 0.793	& 0.928 && 0.794 & 0.843 && 0.798 & 0.847\\
		& 5 && 0.546 & 0.882 && 0.552	& 0.885 && 0.546 & 0.857 && 0.546 & 0.794\\
		& 6 && \multicolumn{1}{c}{$\sbullet$} & \multicolumn{1}{c}{$\sbullet$} && \multicolumn{1}{c}{$\sbullet$} & \multicolumn{1}{c}{$\sbullet$} && \multicolumn{1}{c}{$\sbullet$} & \multicolumn{1}{c}{$\sbullet$} && \multicolumn{1}{c}{$\sbullet$} &  \multicolumn{1}{c}{$\sbullet$} \bigstrut[b]                 \\
		100 & 1 && 0.950 & 0.950 && 0.938	& 0.950 && 0.946 & 0.857 && 0.949 & 0.933\bigstrut[t]\\
		& 2 && 0.944 & 0.955 && 0.945	& 0.944 && 0.941 & 0.824 && 0.938  & 0.870\\
		& 3 && 0.923 & 0.952 && 0.927	& 0.949 &&  0.934 & 0.818 && 0.932 & 0.828\\
		& 4 && 0.848 & 0.936 && 0.849	& 0.935 && 0.845 & 0.832 && 0.845 & 0.834\\
		& 5 && 0.669 & 0.904 && 0.668	& 0.909 && 0.675 & 0.803 && 0.675 &  0.808\\
		& 6 && \multicolumn{1}{c}{$\sbullet$} & \multicolumn{1}{c}{$\sbullet$} && \multicolumn{1}{c}{$\sbullet$} & \multicolumn{1}{c}{$\sbullet$} && \multicolumn{1}{c}{$\sbullet$} & \multicolumn{1}{c}{$\sbullet$} && \multicolumn{1}{c}{$\sbullet$} &  \multicolumn{1}{c}{$\sbullet$} \bigstrut[b]                 \\
		200 & 1 && 0.946 & 0.955 && 0.942	& 0.949 && 0.952 & 0.879 && 0.934 & 0.930 \bigstrut[t]\\
		& 2 && 0.933 & 0.939 && 0.934	& 0.944 && 0.938 & 0.805 && 0.941  &  0.854\\
		& 3 && 0.934 & 0.943 && 0.934	& 0.953 && 0.945 & 0.799 && 0.940 &  0.818\\
		& 4 && 0.883 & 0.948 && 0.886	& 0.946 && 0.885 & 0.850 && 0.884 &  0.860\\
		& 5 && 0.770 & 0.921 && 0.770	& 0.920 && 0.774 & 0.819 && 0.776 &  0.830\\
		& 6 && \multicolumn{1}{c}{$\sbullet$} & \multicolumn{1}{c}{$\sbullet$} && \multicolumn{1}{c}{$\sbullet$} & \multicolumn{1}{c}{$\sbullet$} && \multicolumn{1}{c}{$\sbullet$} & \multicolumn{1}{c}{$\sbullet$} && \multicolumn{1}{c}{$\sbullet$} &  \multicolumn{1}{c}{$\sbullet$} \bigstrut[b]                 \\
		\hline
	\end{tabular}
\end{adjustbox}

\end{table}

\section{\label{sec concl}Conclusion}

This paper has studied identification of an autoregression defined on a
general network, under weak distributional assumptions and without requiring
repeated observations of the network. In this context, identification is
possible for generic parameter values and for generic regressor matrices,
whatever the network. Nevertheless, important cases do exist when
identification fails, either in the original sample space or after some
transformation of the sample space (this could be, for instance, a
transformation aimed at removing fixed effects). We have shown that, in
particular, there are cases where it is impossible to conduct inference that
respects the invariance properties of the model, despite the fact the
parameters may be identifiable from the distribution on the original sample space.

For practical purposes, it may be useful to construct a measure of the
distance from non-identifiability. This goes beyond the scope of the present
paper, but, for example, one may want to have a measure of distance from the
non-identifiability condition $\mathrm{rank}(X,WX)=k$ in Proposition
\ref{lemma identif mean}. One such measure would be the $\left(  k+1\right)
$-th largest singular value of $(X,WX)$, or some norm of the matrix $M_{X}WX$,
possibly upon some normalization of $X$ and $W.$\footnote{To justify these two
measures note that, since $\mathrm{rank}(X)=k,$ (i) $\mathrm{rank}(X,WX)=k$ if
the $\left(  k+1\right)  $-th largest singular value of $(X,WX)$ is zero; (ii)
$\mathrm{rank}(X,WX)=k$ is equivalent to $\operatorname{col}(WX)\subseteq
\operatorname{col}(X)$, or, which is the same, to $M_{X}WX=0$.} Such measures
should help model users to avoid not only the cases in which inference based
on the first moment is impossible, but also cases close to these, in which
inference is likely to be very challenging without additional distributional assumptions.

Finally, it is important to remark that the results in this paper have been
derived under the assumption that the network is fully known and exogenous,
which may be unrealistic in many applications. The study of identification
when the network is (partially) unknown and/or endogenous remains a key
challenge in the literature \citep[e.g.,][]{Blume2015,dePaula2020,Lewbel2019},
and we hope that the results obtained in this paper can prove useful in that
setting too.

\appendix%

\makeatletter\def\@seccntformat#1{Appendix\ \csname the#1\endcsname\quad}%

\makeatother



\section{\label{app further}Further examples when Condition \ref{assum id}
fails}

Further to Examples \ref{exa GI failure} and \ref{exa CBG failure}, other
simple cases in which Condition \ref{assum id} fails are as follows.

\begin{example2}
\label{exa GI incl}Consider the modification of the model in Example
\ref{exa GI} in which \textit{exclusive }averaging is replaced by
\textit{inclusive }averaging, meaning that each unit interacts not only with
all other units in a group but also with itself. If there are $R$ groups, each
of size $m_{r}$, the interaction matrix is $W=\bigoplus_{r=1}^{R}\frac
{1}{m_{r}}\iota_{m_{r}}\iota_{m_{r}}^{\prime}$. Since $\operatorname{col}%
(\bigoplus_{r=1}^{R}\frac{1}{m_{r}}\iota_{m_{r}}\iota_{m_{r}}^{\prime
})=\operatorname{col}(\bigoplus_{r=1}^{R}\iota_{m_{r}})$, Condition
\ref{assum id} is violated (at $\omega=0$) whenever $X$ contains group
intercepts. Thus, in the group interaction model with inclusive averaging and
group fixed effects, Condition \ref{assum id} fails regardless of whether the
model is balanced or not. Recall that, in contrast, with exclusive averaging
Condition \ref{assum id} fails only in the balanced case; see Example
\ref{exa GI failure}. \hfill\qed

\end{example2}

\begin{example2}
\label{exa compl R-bip}Example \ref{exa CBG failure} generalizes immediately
to complete $R$-partite networks, with $R\geq2$ \citep[e.g.,][]{wasserman94}.
Here, and in the next example, $R$ denotes the number of partitions (and not
the number of bipartite networks as in Example \ref{exa netw under cond 1}
(ii)). Such structures are useful, for instance, to model multi-sided markets
in which there are $R$ types of agents, and each agent interacts with all
agents of different type, but with none of the same type. For an
autoregression on a complete $R$-partite network, Condition \ref{assum id} is
violated (at $\omega=0$) whenever $\operatorname{col}(\bigoplus_{r=1}^{R}%
\iota_{m_{r}})\subseteq\operatorname{col}(X)$, where $m_{r}$ denotes the size
of the $r$-th partition, and this is the case if $X$ contains an intercept for
each of the $R$ partitions.\hfill\qed

\end{example2}

\begin{example2}
\label{exa CBG failure app}Examples \ref{exa GI incl} and
\ref{exa compl R-bip} share important similarities, due to the fact that the
networks underlying the two models are \textit{complements} of each other, in
the graph theoretic sense. For both models, in addition to the cases mentioned
in Examples \ref{exa GI incl} and \ref{exa compl R-bip}, the condition
$\operatorname{col}(\bigoplus_{r=1}^{R}\iota_{m_{r}})\subseteq
\operatorname{col}(X)$ leading to a failure of Condition \ref{assum id} is
also satisfied if: (i) $X$ contains an intercept and $R-1$ (linearly
independent) contextual effect terms $Wx_{i}$, for some $x_{1},\ldots
,x_{R-1}\in\mathbb{R}^{n};$ (ii) $X$ contains $R$ (linearly independent)
contextual effect terms $Wx_{1},\ldots,Wx_{R}$, for some $x_{1},\ldots
,x_{R}\in\mathbb{R}^{n}$. Recall that $R$ here denotes the number of groups
for the model of Example \ref{exa GI incl}, and the number of partitions for
the model of Example \ref{exa compl R-bip}, and note that, in order to be full
rank, $X$ can contain at most $R-1$ contextual effect terms if it contains an
intercept, $R$ contextual effect terms otherwise.\hfill\qed

\begin{mycomment}
before I\ had this - check what I\ need to do with it: We have seen in Example
\ref{exa CBG failure} that, for the Complete Bipartite model, Condition
\ref{assum id} fails if $\operatorname{col}(\iota_{p}\bigoplus\iota
_{q})\subseteq\operatorname{col}(X)$. This occurs not only in the case
mentioned there ($X$ contains an intercept for each of the two groups), but
also in the two following circumstances: (i) $X$ contains an intercept and a
contextual effect term $Wx$, for some $x\in\mathbb{R}^{n}$; (ii) $X$ contains
two (linearly independent)\ contextual effect terms $Wx_{1}$ and $Wx_{2}$, for
some $x_{1},x_{2}\in\mathbb{R}^{n}$.\footnote{When $W$ is the interaction
matrix of a complete bipartite model, $Wx\in\operatorname{col}(\iota
_{p}\bigoplus\iota_{q})$, for any $x\in\mathbb{R}^{n}$. Hence, in both cases
(i) and (ii), $\operatorname{col}(\iota_{p}\bigoplus\iota_{q})\subseteq
\operatorname{col}(X)$ is satisfied because $X$ contains two linearly
independent vectors in $\operatorname{col}(\iota_{p}\bigoplus\iota_{q})$%
}
\end{mycomment}

\end{example2}

\begin{mycomment}
	Lemma S.1.2 applies with $\omega=0$, and implies that $\lambda$\symbol{94}\_ML is a constant
\end{mycomment}

\begin{mycomment}
		Note that$\ \bigoplus_{r=1}^{R}\frac{1}{m_{r}}\iota_{m_{r}}\iota_{m_{r}%
		}^{\prime}$ does not have any negative eigenvalues, so some (straightforward)
		modifications would be needed to apply some of the results in the paper to
		this model.
	\end{mycomment}

\section{\label{app lik}The QMLE}

\paragraph{QMLE based on $y$.}

Omitting additive constants, the quasi log-likelihood corresponding to
$\boldsymbol{\varepsilon}\sim\mathrm{N}(0,I_{n})$ in the network
autoregression (\ref{SLM}) is
\begin{equation}
l(\lambda,\beta,\sigma^{2};y)\coloneqq-\frac{n}{2}\log(\sigma^{2}%
)+\log\left\vert \det\left(  S(\lambda)\right)  \right\vert -\frac{1}%
{2\sigma^{2}}(S(\lambda)y-X\beta)^{\prime}(S(\lambda)y-X\beta), \label{loglik}%
\end{equation}
for any $\lambda$ such that $S(\lambda)$ is nonsingular. To avoid tedious
repetitions, we often omit the `quasi-' in front of `log-likelihood'. The QMLE
in most common use maximizes $l(\lambda,\beta,\sigma^{2};y)$ under the
condition that $\lambda$ is in $\Lambda$ (or in a subset thereof), the
parameter space defined in Section \ref{sec id II moment}. That is, the QMLE
of $(\lambda,\beta,\sigma^{2})$ is
\[
(\hat{\lambda}_{\mathrm{ML}},\hat{\beta}_{\mathrm{ML}},\hat{\sigma
}_{\mathrm{ML}}^{2})\coloneqq\underset{\beta\in\mathbb{R}^{k},\hspace
{0.1667em}\sigma^{2}>0,\hspace{0.1667em}\lambda\in\Lambda}{\argmax}%
l(\lambda,\beta,\sigma^{2};y).
\]
Note that if $W$ did not have a negative (resp. positive) eigenvalue, then the
left (resp. right) extreme of $\Lambda$ could be taken to be $-\infty$ (resp.
$+\infty$). Maximization with respect to $\beta$ and $\sigma^{2}$ gives
$\hat{\beta}_{\mathrm{ML}}(\lambda)\coloneqq(X^{\prime}X)^{-1}X^{\prime
}S(\lambda)y$ and $\hat{\sigma}_{\mathrm{ML}}^{2}(\lambda)\coloneqq\frac{1}%
{n}y^{\prime}S^{\prime}(\lambda)M_{X}S(\lambda)y$. Thus, $\hat{\lambda
}_{\mathrm{ML}}$ can be conveniently computed by maximizing over $\Lambda$ the
profile likelihood
\begin{equation}
l(\lambda;y)\coloneqq l(\lambda,\hat{\beta}_{\mathrm{ML}}(\lambda),\hat
{\sigma}_{\mathrm{ML}}^{2}(\lambda))=-\frac{n}{2}\log\left(  \hat{\sigma
}_{\mathrm{ML}}^{2}(\lambda)\right)  +\log\left\vert \det\left(
S(\lambda)\right)  \right\vert , \label{prof lik}%
\end{equation}
where additive constants have again been omitted.

\paragraph{QMLE after reduction by invariance.}

When the model is $\mathcal{G}_{Z}^{1}$-invariant, for some $n\times m$ matrix
$Z$, the principle of invariance advocates reduction to $\mathcal{G}_{Z}^{1}%
$-invariant statistics, that is, statistics that depend on $y$ only through
the maximal invariant $C_{Z}y$. The network model of Example \ref{network FE},
is $\mathcal{G}_{X_{\mathrm{FE}}}^{1}$-invariant, and here we show that, for
this model, reduction by $\mathcal{G}_{X_{\mathrm{FE}}}^{1}$-invariance is
equivalent to the transformation approach proposed by \cite{LeeLiuLin2010} to
remove the network fixed effects. \cite{LeeLiuLin2010} premultiply the model
by the $\left(  n-R\right)  \times n$ matrix $E\coloneqq\bigoplus_{r=1}%
^{R}E_{r}$, where $E_{r}$ is an $m_{r-1}\times m_{r}$ matrix such that
$E_{r}E_{r}^{\prime}=I_{m_{r}-1}$, and $E_{r}^{\prime}E_{r}=M_{\iota_{m_{r}}}$
\citep[in the notation of][$E_{r}=F_{r}^{\prime}$]{LeeLiuLin2010}. Thus
$EE^{\prime}=I_{n-R}$ and $E^{\prime}E=M_{X_{\mathrm{FE}}}$, which shows that
$E$ satisfies the defining properties of a matrix $C_{X_{\mathrm{FE}}}$, and
therefore that premultiplication by $E$ is equivalent to premultiplication by
$C_{X_{\mathrm{FE}}}$. The transformation approach of \cite{LeeLiuLin2010},
which was given a partial likelihood interpretation in that paper, also admits
an invariance interpretation.
\begin{mycomment}
see check-transformation-approach-network-fixed-effects-unbalanced.m for
equivalence with LLL

what happens to QMLE base on Lin 2005 (local diff in Bram, see also LLL page
158) transf? (lik will not be flat , Iguess it behaves more or less like the
QMLE without removal of the FE; started to code this in check-transformation-approach-network-fixed-effects-unbalanced.m)

$\bigoplus_{r=1}^{R}F_{r}^{\prime}$, where $F_{r}$ is an $m_{r}\times m_{r-1}$
matrix such that $F_{r}^{\prime}F_{r}=I_{m_{r}-1}$, and $F_{r}F_{r}^{\prime
}=M_{\iota_{m_{r}}}$.

where the columns of the $m_{r}\times m_{r-1}$ matrix $F_{r}$ are an
orthonormal basis of the orthogonal complement of $\mathrm{col}(\iota_{m_{r}%
})$.

orthonormal basis for $\mathrm{null}(\iota_{m_{r}}^{\prime})$. Note that the
columns of $C_{X_{\mathrm{FE}}}^{\prime}$ are an orthonormal basis of
$\mathrm{null}(\bigoplus_{r=1}^{R}\iota_{m_{r}})$

$F_{r}^{\prime}F_{r}=I_{m_{r}-1}$

$F_{r}F_{r}^{\prime}=M_{\iota_{m_{r}}}$

$F=\bigoplus_{r=1}^{R}F_{r}$, $F^{\prime}F=I_{n-R}$, $FF^{\prime
}=M_{X_{\mathrm{FE}}}$

by def $C_{X_{\mathrm{FE}}}$ is an $\left(  n-R\right)  \times n$ matrix such
that $C_{X_{\mathrm{FE}}}C_{X_{\mathrm{FE}}}^{\prime}=I_{n-R}$ and
$C_{X_{\mathrm{FE}}}^{\prime}C_{X_{\mathrm{FE}}}=M_{X_{\mathrm{FE}}}$

Note that the QMLE based on the Gaussian likelihood of $C_{X_{\mathrm{FE}}%
}\boldsymbol{y}$ is equivalent to the QMLE considered in LLL (partial lik
interpretation there inv justification here)..........

\bigskip%

\begin{equation}
\boldsymbol{y}=\lambda\bigoplus_{r=1}^{R}W_{r}\boldsymbol{y}+\widetilde{X}%
\tilde{\beta}+X_{\mathrm{FE}}\boldsymbol{\alpha}+\sigma\boldsymbol{\varepsilon
},
\end{equation}
$W=\bigoplus_{r=1}^{R}W_{r}$

$X=(\widetilde{X},X_{\mathrm{FE}})$

$\left(  X,WX,W^{2}X\right)  $%

\begin{align}
C_{X_{\mathrm{FE}}}\boldsymbol{y}  &  =\lambda C_{X_{\mathrm{FE}}%
}W\boldsymbol{y}+C_{X_{\mathrm{FE}}}\widetilde{X}\tilde{\beta} +\sigma
C_{X_{\mathrm{FE}}}\boldsymbol{\varepsilon}\\
&  =\lambda W^{\ast}C_{X_{\mathrm{FE}}}\boldsymbol{y}+C_{X_{\mathrm{FE}}%
}\widetilde{X}\tilde{\beta} +\sigma C_{X_{\mathrm{FE}}}\boldsymbol{\varepsilon
}%
\end{align}

$C_{X_{\mathrm{FE}}}W=C_{X_{\mathrm{FE}}}WC_{X_{\mathrm{FE}}}^{\prime
}C_{X_{\mathrm{FE}}}=W^{\ast}C_{X_{\mathrm{FE}}}$

$X_{\mathrm{FE}}$ $n\times R$

$C_{X_{\mathrm{FE}}}\left(  n-R\right)  \times n$

$C_{X_{\mathrm{FE}}}C_{X_{\mathrm{FE}}}^{\prime}=I_{n-R}$

$C_{X_{\mathrm{FE}}}^{\prime}C_{X_{\mathrm{FE}}}=M_{X_{\mathrm{FE}}}$

The instruments are $(C_{X_{\mathrm{FE}}}\widetilde{X},W^{\ast}%
C_{X_{\mathrm{FE}}}\widetilde{X},W^{\ast}{}^{2}C_{X_{\mathrm{FE}}}X)=$
$...........$

$C_{X_{\mathrm{FE}}}(X,WX,W^{2}X)$

\bigskip

$F^{\prime}W=F^{\prime}WFF^{\prime}$

$F^{\prime}W=W^{\ast}F^{\prime}y$

W$^{\ast}$=$F^{\prime}WF$;

$C_{X_{\mathrm{FE}}}W=C_{X_{\mathrm{FE}}}WC_{X_{\mathrm{FE}}}^{\prime}$

$\left(  FW\right)  ^{2}$=$\left(  FWF\right)  ^{2}=$

\bigskip

$F$ $n\times(n-1)$

$F^{\prime}F=I_{n-1}$

$FF^{\prime}=M_{{}}$

\end{mycomment}

\paragraph{Adjusted QMLE.}

When the dimension of $\beta$ is large compared to the sample size, the QMLE
of $(\lambda,\sigma^{2})$ may perform poorly. To tackle this problem, the QMLE
of $(\lambda,\sigma^{2})$ can be adjusted by recentering the profile score
$s(\lambda,\sigma^{2})$ associated to the profile log-likelihood for
$(\lambda,\sigma^{2})$, $l(\lambda,\sigma^{2})\coloneqq l(\hat{\beta
}_{\mathrm{ML}}(\lambda),\sigma^{2},\lambda)$. Under the assumptions
$\mathrm{E}(\boldsymbol{\varepsilon})=0$ and $\mathrm{var}%
(\boldsymbol{\varepsilon})=I_{n}$, the expectation of $s(\lambda,\sigma^{2})$
over the sample space is available analytically and does not depend on the
nuisance parameter $\beta$. Thus, calculation of the adjusted profile score
$s_{\mathrm{a}}(\lambda,\sigma^{2})\coloneqq s(\sigma^{2},\lambda
)-\mathrm{E}(s(\lambda,\sigma^{2}))$ is straightforward. Given $s_{\mathrm{a}%
}(\lambda,\sigma^{2})$, one can define the adjusted likelihood $l_{\mathrm{a}%
}(\lambda,\sigma^{2})$ as the function with gradient equal to $s_{\mathrm{a}%
}(\lambda,\sigma^{2})$, and hence the adjusted QMLE $(\hat{\lambda
}_{\mathrm{aML}},\hat{\sigma}_{\mathrm{aML}}^{2})$ as the maximizer of
$l_{\mathrm{a}}(\lambda,\sigma^{2})$. Also, letting $\hat{\sigma
}_{\mathrm{aML}}^{2}(\lambda)$ be the adjusted QMLE of $\sigma^{2}$ for given
$\lambda$, we define the adjusted likelihood for $\lambda$ only as
$l_{\mathrm{a}}(\lambda)\coloneqq l_{\mathrm{a}}(\lambda,\hat{\sigma
}_{\mathrm{aML}}^{2}(\lambda))$. See \cite{Yu2015} for details on these
constructions. By standard arguments
\citep[available for instance in][]{RahmanKing1997}, $l_{\mathrm{a}}%
(\lambda,\sigma^{2})$ corresponds to the density of the maximal invariant
$C_{X}\boldsymbol{y} $ under $\mathcal{G}^{1}_{X}$, for any network
autoregression model violating Condition \ref{assum id} and for any network
error model. \begin{mycomment}
	 By standard arguments
	\citep[available for instance in][]{RahmanKing1997}, $l_{\mathrm{a}}(\lambda)$
	corresponds to the density of the maximal invariant $C_{X}\boldsymbol{y}/\left\Vert C_{X}\boldsymbol{y}\right\Vert $
	under $\mathcal{G}_{X}$, for any network autoregression model violating
	Condition \ref{assum id} and for any network error model. Then, the flatness
	of $l_{\mathrm{a}}(\lambda)$ can be understood in terms of the distribution of
	$v(\boldsymbol{y})$ being free of $\lambda$ if the distribution of
	$\boldsymbol{\varepsilon}$ is free of $\lambda$, which follows from equation \eqref{Cy eq}.
	
\end{mycomment}

\section{\label{sec proofs}Proofs}

\begin{lemma}
\label{lemma same prof lik}The network autoregression (\ref{SLM}) and the
network error model (\ref{SEM}) imply the same profile quasi log-likelihood
function for $(\lambda,\sigma^{2})$ if and only if $\mathrm{rank}(X,WX)=k$.
\end{lemma}

\begin{pff}
[Proof of Lemma \ref{lemma same prof lik}]On concentrating the nuisance
parameter $\beta$ out of the likelihood (\ref{loglik}), the profile quasi
log-likelihood for $(\lambda,\sigma^{2})$ in a network autoregression is, up
to an additive constant,%
\begin{equation}
l(\lambda,\sigma^{2})\coloneqq l(\hat{\beta}_{\mathrm{ML}}(\lambda),\sigma
^{2},\lambda)=-\frac{n}{2}\log(\sigma^{2})+\log\left\vert \det\left(
S(\lambda)\right)  \right\vert -\frac{1}{2\sigma^{2}}y^{\prime}S^{\prime
}(\lambda)M_{X}S(\lambda)y. \label{lik sig lam SAR}%
\end{equation}
Similarly, the profile quasi log-likelihood function for $(\lambda,\sigma
^{2})$ in a network error model, based again on the assumption
$\boldsymbol{\varepsilon} \sim\mathrm{N}(0,I_{n})$, is
\begin{equation}
l(\lambda,\sigma^{2})\coloneqq-\frac{n}{2}\log(\sigma^{2})+\log\left\vert
\det\left(  S(\lambda)\right)  \right\vert -\frac{1}{2\sigma^{2}}y^{\prime
}S^{\prime}(\lambda)M_{S(\lambda)X}S(\lambda)y. \label{lik sig lambda}%
\end{equation}
The two log-likelihood functions are the same if and only if $M_{S(\lambda
)X}=M_{X}$ for any $\lambda$ such that $S(\lambda)$ is invertible. But, for
any $\lambda$ such that $S(\lambda)$ is invertible, the condition
$M_{S(\lambda)X}=M_{X}$ is equivalent to $\operatorname{col}(S(\lambda
)X)=\operatorname{col}(X)$, and hence to $\operatorname{col}(WX)\subseteq
\operatorname{col}(X)$, which in turn is the same as $\mathrm{rank}(X,WX)=k.$
\end{pff}

\begin{pff}
[Proof of Proposition \ref{lemma identif mean}]The parameter $(\lambda,\beta)$
is identified on $\Lambda_{\mathrm{u}}\times\mathbb{R}^{k}$ from
$\mathrm{E}(\boldsymbol{y})=S^{-1}(\lambda)X\beta$ if $S^{-1}(\overset{\sim
}{\smash{\lambda}\rule{0pt}{1.13ex}})X\overset{\sim
}{\smash{\beta}\rule{0pt}{1.13ex}}=S^{-1}(\overset{\sim
}{\smash{\lambdatilde}\rule{0pt}{1.5ex}})X\overset{\sim
}{\smash{\betatilde}\rule{0pt}{1.5ex}}$ implies $( \overset{\sim
}{\smash{\lambda}\rule{0pt}{1.13ex}},\overset{\sim
}{\smash{\beta}\rule{0pt}{1.13ex}}) =(\overset{\sim
}{\smash{\lambdatilde}\rule{0pt}{1.5ex}},\overset{\sim
}{\smash{\betatilde}\rule{0pt}{1.5ex}})$ for any two values $( \overset{\sim
}{\smash{\lambda}\rule{0pt}{1.13ex}},\overset{\sim
}{\smash{\beta}\rule{0pt}{1.13ex}}) ,(\overset{\sim
}{\smash{\lambdatilde}\rule{0pt}{1.5ex}},\overset{\sim
}{\smash{\betatilde}\rule{0pt}{1.5ex}})$ of $(\lambda,\beta)$ in
$\Lambda_{\mathrm{u}}\times\mathbb{R}^{k}.$ One immediately has that
$S^{-1}(\overset{\sim}{\smash{\lambda}\rule{0pt}{1.13ex}})X\overset{\sim
}{\smash{\beta}\rule{0pt}{1.13ex}}=S^{-1}(\overset{\sim
}{\smash{\lambdatilde}\rule{0pt}{1.5ex}})X\overset{\sim
}{\smash{\betatilde}\rule{0pt}{1.5ex}}$ if and only if%

\begin{equation}
X(\overset{\sim}{\smash{\beta}\rule{0pt}{1.13ex}}-\overset{\sim
}{\smash{\betatilde}\rule{0pt}{1.5ex}})+WX(\overset{\sim
}{\smash{\lambda}\rule{0pt}{1.13ex}}\overset{\sim
}{\smash{\betatilde}\rule{0pt}{1.5ex}}-\overset{\sim
}{\smash{\lambdatilde}\rule{0pt}{1.5ex}}\overset{\sim
}{\smash{\beta}\rule{0pt}{1.13ex}})=0. \label{XWX}%
\end{equation}
We analyze separately three (exhaustive) cases, depending on the rank of the
$n\times2k$ matrix $(X,WX)$. Recall that $X$ is assumed to be of full column rank.

\begin{enumerate}
\item[(a)] $\mathrm{rank}(X,WX)=2k$. In this case equation (\ref{XWX}) is
equivalent to $\overset{\sim}{\smash{\beta}\rule{0pt}{1.13ex}}=\overset{\sim
}{\smash{\betatilde}\rule{0pt}{1.5ex}}$ and $\overset{\sim
}{\smash{\lambda}\rule{0pt}{1.13ex}}\overset{\sim
}{\smash{\betatilde}\rule{0pt}{1.5ex}}=\overset{\sim
}{\smash{\lambdatilde}\rule{0pt}{1.5ex}}\overset{\sim
}{\smash{\beta}\rule{0pt}{1.13ex}},$ from which $(\overset{\sim
}{\smash{\lambda}\rule{0pt}{1.13ex}},\overset{\sim
}{\smash{\beta}\rule{0pt}{1.13ex}}) =(\overset{\sim
}{\smash{\lambdatilde}\rule{0pt}{1.5ex}},\overset{\sim
}{\smash{\betatilde}\rule{0pt}{1.5ex}})$ if and only if $\overset{\sim
}{\smash{\beta}\rule{0pt}{1.13ex}}=\overset{\sim
}{\smash{\betatilde}\rule{0pt}{1.5ex}}\neq0.$ That is, $\left(  \lambda
,\beta\right)  $ is identified on $\Lambda_{\mathrm{u}}\times\mathbb{R}%
^{k}\backslash\{0\}$ from $\mathrm{E}(\boldsymbol{y}).$

\item[(b)] $k<\mathrm{rank}(X,WX)<2k$. Partition $X$ as $(X_{1},X_{2})$ where
$X_{1}$ is $n\times k_{1}$ and $X_{2}$ is $n\times k_{2}$, with $0<k_{1}<k$.
The case $k<\mathrm{rank}(X,WX)<2k$ may be characterized by assuming
$\mathrm{rank}(X,WX_{1})=k+k_{1}$ and $WX_{2}=XB+WX_{1}C,$ for some $k\times
k_{2}$ matrix $B$ and some $k_{1}\times k_{2}$ matrix $C$, so that
$\mathrm{rank}(X,WX)=k+k_{1}$. Replacing $WX$ with $\left(  WX_{1}%
,XB+WX_{1}C\right)  $ in (\ref{XWX}), and letting $\left(  \beta_{1}^{\prime
},\beta_{2}^{\prime}\right)  ^{\prime} $ be the partition of $\beta^{\prime}$
conformable with that of $X$, we obtain
\[
X(\overset{\sim}{\smash{\beta}\rule{0pt}{1.13ex}}-\overset{\sim
}{\smash{\betatilde}\rule{0pt}{1.5ex}}+B(\overset{\sim
}{\smash{\lambda}\rule{0pt}{1.13ex}}\overset{\sim
}{\smash{\betatilde}\rule{0pt}{1.5ex}}_{2}-\overset{\sim
}{\smash{\lambdatilde}\rule{0pt}{1.5ex}}\overset{\sim
}{\smash{\beta}\rule{0pt}{1.13ex}} _{2}))+WX_{1}(\overset{\sim
}{\smash{\lambda}\rule{0pt}{1.13ex}}\overset{\sim
}{\smash{\betatilde}\rule{0pt}{1.5ex}}_{1}-\overset{\sim
}{\smash{\lambdatilde}\rule{0pt}{1.5ex}}\overset{\sim
}{\smash{\beta}\rule{0pt}{1.13ex}}_{1}+C(\overset{\sim
}{\smash{\lambda}\rule{0pt}{1.13ex}} \overset{\sim
}{\smash{\betatilde}\rule{0pt}{1.5ex}}_{2}-\overset{\sim
}{\smash{\lambdatilde}\rule{0pt}{1.5ex}}\overset{\sim
}{\smash{\beta}\rule{0pt}{1.13ex}}_{2}))=0,
\]
which is satisfied if and only if $\overset{\sim
}{\smash{\beta}\rule{0pt}{1.13ex}}-\overset{\sim
}{\smash{\betatilde}\rule{0pt}{1.5ex}}+B(\overset{\sim
}{\smash{\lambda}\rule{0pt}{1.13ex}}\overset{\sim
}{\smash{\betatilde}\rule{0pt}{1.5ex}}_{2}-\overset{\sim
}{\smash{\lambdatilde}\rule{0pt}{1.5ex}}\overset{\sim
}{\smash{\beta}\rule{0pt}{1.13ex}}_{2})=0$ and $\overset{\sim
}{\smash{\lambda}\rule{0pt}{1.13ex}}\overset{\sim
}{\smash{\betatilde}\rule{0pt}{1.5ex}}_{1}-\overset{\sim
}{\smash{\lambdatilde}\rule{0pt}{1.5ex}}\overset{\sim
}{\smash{\beta}\rule{0pt}{1.13ex}}_{1}+C(\overset{\sim
}{\smash{\lambda}\rule{0pt}{1.13ex}}\overset{\sim
}{\smash{\betatilde}\rule{0pt}{1.5ex}}_{2}-\overset{\sim
}{\smash{\lambdatilde}\rule{0pt}{1.5ex}}\overset{\sim
}{\smash{\beta}\rule{0pt}{1.13ex}} _{2})=0$. As a linear system in the
unknowns $\overset{\sim}{\smash{\lambdatilde}\rule{0pt}{1.5ex}}$ and
$\overset{\sim}{\smash{\betatilde}\rule{0pt}{1.5ex}}$, these two equations
are
\begin{equation}
M(\overset{\sim}{\smash{\lambda}\rule{0pt}{1.13ex}},\overset{\sim
}{\smash{\beta}\rule{0pt}{1.13ex}})%
\begin{pmatrix}
\overset{\sim}{\smash{\lambdatilde}\rule{0pt}{1.5ex}}\\
\overset{\sim}{\smash{\betatilde}\rule{0pt}{1.5ex}}%
\end{pmatrix}
=%
\begin{pmatrix}
\overset{\sim}{\smash{\beta}\rule{0pt}{1.13ex}}\\
0_{k_{1}}%
\end{pmatrix}
, \label{system}%
\end{equation}
where the matrix
\[
M(\overset{\sim}{\smash{\lambda}\rule{0pt}{1.13ex}},\overset{\sim
}{\smash{\beta}\rule{0pt}{1.13ex}})\coloneqq\left(
\begin{array}
[c]{cc}%
B\overset{\sim}{\smash{\beta}\rule{0pt}{1.13ex}}_{2} & I_{k}-\overset{\sim
}{\smash{\lambda}\rule{0pt}{1.13ex}}(0_{k,k_{1}},B)\\
\overset{\sim}{\smash{\beta}\rule{0pt}{1.13ex}}_{1}+C\overset{\sim
}{\smash{\beta}\rule{0pt}{1.13ex}}_{2} & -\overset{\sim
}{\smash{\lambda}\rule{0pt}{1.13ex}}(I_{k_{1}},C)
\end{array}
\right)
\]
is of dimension $\left(  k+k_{1}\right)  \times(1+k)$. Now, identification of
$(\lambda,\beta)$ from $\mathrm{E}(\boldsymbol{y})$ is equivalent to
$(\overset{\sim}{\smash{\lambda}\rule{0pt}{1.13ex}},\overset{\sim
}{\smash{\beta}\rule{0pt}{1.13ex}})$ being the unique solution to system
(\ref{system}), and this occurs if and only if ${\mathrm{rank}(M(\overset{\sim
}{\smash{\lambda}\rule{0pt}{1.13ex}},\overset{\sim
}{\smash{\beta}\rule{0pt}{1.13ex}}))}=1+k$, or, equivalently, $\det
(M(\overset{\sim}{\smash{\lambda}\rule{0pt}{1.13ex}},\overset{\sim
}{\smash{\beta}\rule{0pt}{1.13ex}})^{\prime}M(\overset{\sim
}{\smash{\lambda}\rule{0pt}{1.13ex}},\overset{\sim
}{\smash{\beta}\rule{0pt}{1.13ex}}))\neq0.$ But ${\det(M(\overset{\sim
}{\smash{\lambda}\rule{0pt}{1.13ex}},\overset{\sim
}{\smash{\beta}\rule{0pt}{1.13ex}})^{\prime}M(\overset{\sim
}{\smash{\lambda}\rule{0pt}{1.13ex}},\overset{\sim
}{\smash{\beta}\rule{0pt}{1.13ex}}))}$ is a polynomial in $(\overset{\sim
}{\smash{\lambda}\rule{0pt}{1.13ex}},\overset{\sim
}{\smash{\beta}\rule{0pt}{1.13ex}})$ and hence the set of its zeros is either
the whole $\mathbb{R}^{k+1}$ or has zero measure with respect to
$\mu_{\mathbb{R}^{k+1}}$. The former case is easily ruled out (e.g.,
$M(\overset{\sim}{\smash{\lambda}\rule{0pt}{1.13ex}},\overset{\sim
}{\smash{\beta}\rule{0pt}{1.13ex}})$ has rank $k+1$ for $(\overset{\sim
}{\smash{\lambda}\rule{0pt}{1.13ex}},\overset{\sim
}{\smash{\beta}\rule{0pt}{1.13ex}})=(0,(1_{k_{1}}^{\prime},0_{k_{2}}^{\prime
})^{\prime})$), which means that $(\lambda,\beta)$ is generically identified
from $\mathrm{E}(\boldsymbol{y})$.

\begin{mycomment}
	example of value of $\left(  \lambda,\beta\right)  $ that is identified:\ for
$\left(  \lambda,\beta\right)  =\left(  0,%
\begin{array}
[c]{c}%
1_{k_{1}}\\
0_{k_{2}}%
\end{array}
\right)  ,$%
\[
M_{\lambda,\beta}=\left(
\begin{array}
[c]{cc}%
0_{k} & I_{k}\\
1_{k_{1}} & 0_{k_{1},k}%
\end{array}
\right)  ,
\]
which has full col rank
example of value of $\left(  \lambda,\beta\right)  $ that is not identified:%
\[
M_{\lambda,0_{k}}\coloneqq\left(
\begin{array}
[c]{cc}%
0_{k} & I_{k}-\lambda(0_{k,k_{1}},B)\\
0_{k_{1}} & -\lambda(I_{k_{1}},C)
\end{array}
\right)  ,\text{ }\mathrm{rank}(M_{\lambda,0_{k}})<1+k\text{ }%
\]
\end{mycomment}

\begin{mycomment}
in the supplement give the unidentified set and perhaps the particular case in proof\_lemma\_6.1.tex
\end{mycomment}

\item[(c)] $\mathrm{rank}(X,WX)=k$. This happens if and only if there is a
$k\times k$ matrix $A$ such that $WX=XA$. In that case, equation (\ref{XWX})
becomes $X(\overset{\sim}{\smash{\beta}\rule{0pt}{1.13ex}}-\overset{\sim
}{\smash{\betatilde}\rule{0pt}{1.5ex}}+A(\overset{\sim
}{\smash{\lambda}\rule{0pt}{1.13ex}}\overset{\sim
}{\smash{\betatilde}\rule{0pt}{1.5ex}}-\overset{\sim
}{\smash{\lambdatilde}\rule{0pt}{1.5ex}}\overset{\sim
}{\smash{\beta}\rule{0pt}{1.13ex}}))=0$, which, since $\mathrm{rank}(X)=k$, is
equivalent to $\overset{\sim}{\smash{\beta}\rule{0pt}{1.13ex}} -\overset{\sim
}{\smash{\betatilde}\rule{0pt}{1.5ex}}+A(\overset{\sim
}{\smash{\lambda}\rule{0pt}{1.13ex}}\overset{\sim
}{\smash{\betatilde}\rule{0pt}{1.5ex}}-\overset{\sim
}{\smash{\lambdatilde}\rule{0pt}{1.5ex}}\overset{\sim
}{\smash{\beta}\rule{0pt}{1.13ex}} )=0$. Rewrite the last equality as
$(I_{k}-\overset{\sim}{\smash{\lambdatilde}\rule{0pt}{1.5ex}} A)\overset{\sim
}{\smash{\beta}\rule{0pt}{1.13ex}}-(I_{k}-\overset{\sim
}{\smash{\lambda}\rule{0pt}{1.13ex}} A)\overset{\sim
}{\smash{\betatilde}\rule{0pt}{1.5ex}}=0$. Since the eigenvalues of $A$ are
eigenvalues of $W,$ $I_{k}-\lambda A$ is invertible for any $\lambda\in
\Lambda_{\mathrm{u}}$, and therefore $\overset{\sim
}{\smash{\beta}\rule{0pt}{1.13ex}}=(I_{k}-\overset{\sim
}{\smash{\lambdatilde}\rule{0pt}{1.5ex}} A)^{-1}(I_{k}-\overset{\sim
}{\smash{\lambda}\rule{0pt}{1.13ex}} A)\overset{\sim
}{\smash{\betatilde}\rule{0pt}{1.5ex}}$. This shows that for any
$(\overset{\sim}{\smash{\lambdatilde}\rule{0pt}{1.5ex}},\overset{\sim
}{\smash{\betatilde}\rule{0pt}{1.5ex}} )\in\Lambda_{\mathrm{u}}\times
\mathbb{R}^{k}$, it is possible to find $( \overset{\sim
}{\smash{\lambda}\rule{0pt}{1.13ex}},\overset{\sim
}{\smash{\beta}\rule{0pt}{1.13ex}}) \neq(\overset{\sim
}{\smash{\lambdatilde}\rule{0pt}{1.5ex}},\overset{\sim
}{\smash{\betatilde}\rule{0pt}{1.5ex}})$ such that $S^{-1}(\overset{\sim
}{\smash{\lambda}\rule{0pt}{1.13ex}})X\overset{\sim
}{\smash{\beta}\rule{0pt}{1.13ex}} =S^{-1}(\overset{\sim
}{\smash{\lambdatilde}\rule{0pt}{1.5ex}})X\overset{\sim
}{\smash{\betatilde}\rule{0pt}{1.5ex}}$.
\end{enumerate}

Summarizing, $\left(  \lambda,\beta\right)  $ is generically identified from
$\mathrm{E}(\boldsymbol{y})$, and hence generically identified, on
$\Lambda_{\mathrm{u}}\times\mathbb{R}^{k}$ in cases (a) and (b), and not
identified from $\mathrm{E}(\boldsymbol{y})$ on $\Lambda_{\mathrm{u}}%
\times\mathbb{R}^{k}$ in case (c).
\end{pff}

\begin{mycomment}
		check that $(\lambda,\beta)$ is a solution:%
		\begin{equation}
		\left[
		\begin{array}
		[c]{cc}%
		-B\beta_{2} & I_{k}+\lambda(0,B)\\
		-\left(  \beta_{1}+C\beta_{2}\right)   & \lambda(I_{k_{1}},C)
		\end{array}
		\right]  \binom{\lambda}{\beta}=\binom{-\lambda B\beta_{2}+\beta
			+\lambda(0_{k,k_{1}},B)\underset{k\times1}{\beta}}{-\lambda\beta_{1}-\lambda
			C\beta_{2}+\lambda(I_{k_{1}},C)\beta}=\binom{-\lambda B\beta_{2}+\beta+\lambda
			B\beta_{2}}{-\lambda\beta_{1}-\lambda C\beta_{2}+\lambda\beta_{1}+\lambda
			C\beta_{2})}=\binom{\beta}{0}.
		\end{equation}
	\end{mycomment}

\begin{pff}
[Proof of Proposition \ref{lemma identif var}]This proof is similar to that of
Lemma 4.2 in \cite{Preinerstorfer2015}. Under the assumption that
$\mathrm{var}(\boldsymbol{\varepsilon})=I_{n}$, $\mathrm{var}(\boldsymbol{y}%
)=\sigma^{2}(S^{\prime}(\lambda)S(\lambda))^{-1}$. We want to establish that,
if $\overset{\sim}{\smash{\sigmatilde}\rule{0pt}{1ex}}\rule{0mm}{1.13ex}%
^{2}S^{\prime}(\overset{\sim}{\smash{\lambda}\rule{0pt}{1.13ex}}%
)S(\overset{\sim}{\smash{\lambda}\rule{0pt}{1.13ex}})=\overset{\sim
}{\smash{\sigma}\rule{0pt}{0.63ex}}\rule{0mm}{1.13ex}^{2}S^{\prime
}(\overset{\sim}{\smash{\lambdatilde}\rule{0pt}{1.5ex}})S(\overset{\sim
}{\smash{\lambdatilde}\rule{0pt}{1.5ex}})$ for any two parameter values
${(\overset{\sim}{\smash{\lambda}\rule{0pt}{1.13ex}},\overset{\sim
}{\smash{\sigma}\rule{0pt}{0.63ex}}\rule{0mm}{1.13ex}^{2}),(\overset{\sim
}{\smash{\lambdatilde}\rule{0pt}{1.5ex}},\overset{\sim
}{\smash{\sigmatilde}\rule{0pt}{1ex}}\rule{0mm}{1.13ex}^{2})}\in\Lambda
\times(0,\infty)$, then $(\overset{\sim}{\smash{\lambda}\rule{0pt}{1.13ex}%
},\overset{\sim}{\smash{\sigma}\rule{0pt}{0.63ex}}\rule{0mm}{1.13ex}%
^{2})=(\overset{\sim}{\smash{\lambdatilde}\rule{0pt}{1.5ex}},\overset{\sim
}{\smash{\sigmatilde}\rule{0pt}{1ex}}\rule{0mm}{1.13ex}^{2})$. The maintained
assumption that $W$ has at least one negative eigenvalue and at least one
positive eigenvalue guarantees the existence of a nonzero vector
$f\in\mathrm{null}(W-I_{n})$ and a nonzero vector $g\in\mathrm{null}%
(W-\omega_{\min}I_{n})$. Multiplying both sides of the equality $\overset{\sim
}{\smash{\sigmatilde}\rule{0pt}{1ex}}\rule{0mm}{1.13ex}^{2}S^{\prime
}(\overset{\sim}{\smash{\lambda}\rule{0pt}{1.13ex}})S(\overset{\sim
}{\smash{\lambda}\rule{0pt}{1.13ex}})=\overset{\sim
}{\smash{\sigma}\rule{0pt}{0.63ex}}\rule{0mm}{1.13ex}^{2}S^{\prime
}(\overset{\sim}{\smash{\lambdatilde}\rule{0pt}{1.5ex}})S(\overset{\sim
}{\smash{\lambdatilde}\rule{0pt}{1.5ex}})$ by $f^{\prime}$ on the left and $f$
on the right gives $\overset{\sim}{\smash{\sigmatilde}\rule{0pt}{1ex}%
}\rule{0mm}{1.13ex}^{2}(1-\overset{\sim}{\smash{\lambda}\rule{0pt}{1.13ex}%
})^{2}f^{\prime}f=\overset{\sim}{\smash{\sigma}\rule{0pt}{0.63ex}%
}\rule{0mm}{1.13ex}^{2}(1-\overset{\sim}{\smash{\lambdatilde}\rule{0pt}{1.5ex}%
})^{2}f^{\prime}f$. Since $1-\lambda>0$ for any $\lambda\in\Lambda$, and
$f^{\prime}f\neq0,$ the last equality is equivalent to $\overset{\sim
}{\smash{\sigmatilde}\rule{0pt}{1ex}}/\overset{\sim
}{\smash{\sigma}\rule{0pt}{0.63ex}}=(1-\overset{\sim
}{\smash{\lambdatilde}\rule{0pt}{1.5ex}})/(1-\overset{\sim
}{\smash{\lambda}\rule{0pt}{1.13ex}})$. Repeating with $g$ in place of $f$
gives $\overset{\sim}{\smash{\sigmatilde}\rule{0pt}{1ex}}/\overset{\sim
}{\smash{\sigma}\rule{0pt}{0.63ex}}={(1-\overset{\sim
}{\smash{\lambdatilde}\rule{0pt}{1.5ex}}\omega_{\min})}/(1-\overset{\sim
}{\smash{\lambda}\rule{0pt}{1.13ex}}\omega_{\min})$. Thus, we must have
$(1-\overset{\sim}{\smash{\lambda}\rule{0pt}{1.13ex}}\omega_{\min
})/(1-\overset{\sim}{\smash{\lambda}\rule{0pt}{1.13ex}})=(1-\overset{\sim
}{\smash{\lambdatilde}\rule{0pt}{1.5ex}}\omega_{\min})/(1-\overset{\sim
}{\smash{\lambdatilde}\rule{0pt}{1.5ex}})$. Since the function $\lambda
\mapsto\left(  1-\lambda\omega_{\min}\right)  /\left(  1-\lambda\right)  $ is
strictly increasing on $\Lambda$, we have $\overset{\sim
}{\smash{\lambda}\rule{0pt}{1.13ex}}=\overset{\sim
}{\smash{\lambdatilde}\rule{0pt}{1.5ex}}$, and hence $\overset{\sim
}{\smash{\sigma}\rule{0pt}{0.63ex}}\rule{0mm}{1.13ex}=\overset{\sim
}{\smash{\sigmatilde}\rule{0pt}{1ex}}\rule{0mm}{1.13ex}$.
\end{pff}

\begin{mycomment}
	could do a slightly more general proof with two arbitrary distinct eigenvalues
	$\omega_{1}$ $\omega_{2}$ but I'm interested in identif over $\Lambda$...
\end{mycomment}

\begin{pff}
[Proof of Lemma \ref{lemma model inv}]For any $\lambda$ such that $S(\lambda)$
is nonsingular and under Assumption \ref{assum distrib}, it is clear from the
reduced form $\boldsymbol{y}=S^{-1}(\lambda)X\beta+\sigma S^{-1}%
(\lambda)\boldsymbol{\varepsilon}$ that a network autoregression is invariant
under $\mathcal{G}_{X}$ if and only if $\operatorname{col}(S^{-1}%
(\lambda)X)=\operatorname{col}(X)$, or, which is the same, $\operatorname{col}%
(S(\lambda)X)=\operatorname{col}(X)$. As noted in the proof of Lemma
\ref{lemma same prof lik}, the condition $\operatorname{col}(S(\lambda
)X)=\operatorname{col}(X)$ for any $\lambda$ such that $S(\lambda)$ is
invertible is equivalent to $\mathrm{rank}(X,WX)=k$.
\end{pff}

\begin{pff}
[Proof of Proposition \ref{prop prof lik Assumption A}]Since $C_{Z} Z=0$ for
any full column rank matrix $Z$, it follows that $l(\lambda,\beta,\sigma
^{2};C_{X}y)$ does not depend on $\beta$, and $l(\lambda,\beta,\sigma
^{2};C_{X_{\omega}}y)$ does not depend on the component of $\beta$ associated
to $X_{\omega}.$ The more general statement (i) follows from Theorem
\ref{theo non-identif}. Let us move to part (ii). For any $\lambda$ such that
$\mathrm{rank}\left(  S(\lambda)\right)  =n$, and for any $y\notin%
\mathrm{null}(M_{X}S(\lambda))$, the profile log-likelihood $l(\lambda;y)$ for
a network autoregression is given by equation (\ref{prof lik}). Note that
equation (\ref{prof lik}) holds a.s.\ for any fixed $\lambda$ such that
\textrm{rank}$\left(  S(\lambda)\right)  =n$, because $\mathrm{null}%
(M_{X}S(\lambda))$ is a $\mu_{\mathbb{R}^{n}}$-null set when \textrm{rank}%
$\left(  S(\lambda)\right)  =n$ (since $k<n$). If Condition \ref{assum id} is
violated for an eigenvalue $\omega$ of $W$, then $M_{X}(\omega I_{n}-W)=0$ and
hence $M_{X}S(\lambda)=(1-\lambda\omega)M_{X}$, which substituted into
(\ref{prof lik}) gives
\begin{equation}
l(\lambda;y)=\log\left\vert \det\left(  S(\lambda)\right)  \right\vert
-n\log\left\vert 1-\lambda\omega\right\vert -\frac{n}{2}\log(y^{\prime}%
M_{X}y), \label{prof lik ASSUM1 violated}%
\end{equation}
for any $y\notin\operatorname{col}(X)$. Since a violation of Condition
\ref{assum id} implies $\mathrm{rank}(X,WX)=k$, equation
(\ref{prof lik ASSUM1 violated}) also applies to a network error model, by
Lemma \ref{lemma same prof lik}. Part (ii) of the proposition follows on
noting that the terms in (\ref{prof lik ASSUM1 violated}) that contain
$\lambda$ do not contain $y$. \begin{mycomment}
Next, let $s(\lambda)$ be the profile score associated with $l(\lambda)$, let
$s_{\mathrm{a}}(\lambda)\coloneqq s(\lambda)-\mathrm{E}(s(\lambda))$ be its
adjusted counterpart, and let $l_{\mathrm{a}}^{\ast}(\lambda)\coloneqq\int
s_{\mathrm{a}}(\lambda)\diff\lambda$ be the likelihood corresponding to
$s_{\mathrm{a}}(\lambda)$. It can be easily verified that $l_{\mathrm{a}%
}(\lambda)=\frac{n-k}{n}l_{\mathrm{a}}^{\ast}(\lambda)$ (the adjusted profile
likelihood $l_{\mathrm{a}}$ being defined in Appendix \ref{app lik}). If
Condition \ref{assum id} is violated, then, from part (i), $\mathrm{E}\left(
s(\lambda)\right)  =s(\lambda)$, and hence $s_{\mathrm{a}}(\lambda)=0$, which
in turn implies that $l_{\mathrm{a}}^{\ast}(\lambda)$, and hence
$l_{\mathrm{a}}(\lambda)$, is constant. This completes the proof.
\end{mycomment}

\end{pff}

\begingroup\setlength{\bibsep}{6pt} \setstretch{1.15}
\bibliographystyle{elsart-harv}
\bibliography{spatial_identif}
\endgroup

\end{document}